\documentclass[aps,prb,twocolumn,superscriptaddress,citeautoscript]{revtex4-1}
\usepackage{graphicx}
\usepackage{mathrsfs}
\usepackage{amsmath}
\usepackage{amssymb}
\usepackage{color}
\usepackage[nice]{nicefrac}
\usepackage{float}
\usepackage{lipsum}
\usepackage[a4paper,lmargin=2cm,rmargin=1.2cm,tmargin=2.2cm,bmargin=2.2cm]{geometry}

\usepackage{color}
\definecolor{darkred}{rgb}{0.6,0.,0.}
\definecolor{darkgreen}{rgb}{0.,0.5,0.}
\definecolor{darkblue}{rgb}{0.,0.,0.6}

\usepackage[
breaklinks,
colorlinks=true,
linkcolor=darkred,
citecolor=darkgreen,
urlcolor=darkblue]{hyperref}

\def\be{\begin{eqnarray}}
\def\ee{\end{eqnarray}}
\def\nn{\nonumber}

\def\comment#1{(see comment in source)}
\def\state#1{|#1\rangle}

\def\spinu{|\uparrow\rangle}
\def\spind{|\downarrow\rangle}

\def\Pf{{\rm Pf}}

\newif\ifcomments
\commentsfalse

\newif\ifoldpara
\oldparafalse

\begin{document}

\title{Josephson Coupled Moore-Read States}

\author{Gunnar M\"oller}
\affiliation{
TCM Group, Cavendish Laboratory, University of Cambridge, Cambridge CB3 0HE, United Kingdom}
\author{Layla Hormozi}
\affiliation{Department of Mathematical Physics, National University of Ireland, Maynooth, Ireland}
\author{Joost Slingerland}
\affiliation{Department of Mathematical Physics, National University of Ireland, Maynooth, Ireland}
\affiliation{Dublin Institute for Advanced Studies, School of Theoretical Physics, 10 Burlington Rd, Dublin, Ireland}
\author{Steven H. Simon}
\affiliation{Rudolf Peierls Centre for Theoretical Physics, University of Oxford, Oxford OX1 3NP, United Kingdom}

\date{September 22, 2014}

\begin{abstract}
We study a quantum Hall bilayer system of bosons at total filling factor $\nu = 1$, and study the phase that results from short ranged pair-tunneling combined with short ranged interlayer interactions. 
We introduce two exactly solvable model Hamiltonians which both yield the coupled Moore-Read state [Phys.~Rev.~Lett.~{\bf 108}, 256809 (2012)] as a ground state, when projected onto fixed particle numbers in each layer. One of these Hamiltonians describes a gapped topological phase while the other is gapless. However, on introduction of a pair tunneling term, the second system becomes gapped and develops the same topological order as the gapped Hamiltonian. Supported by the exact solution of the full zero-energy quasihole spectrum and a conformal field theory approach, we develop an intuitive picture of this system as two coupled composite fermion superconductors. In this language, pair tunneling provides a Josephson coupling of the superconducting phases of the two layers, and gaps out the Goldstone mode associated with particle transport between the layers. In particular, this implies that quasiparticles are confined between the layers. In the bulk, the resulting phase has the topological order of the Halperin 220 phase with U(1)$_2 \times$ U(1)$_2$ topological order, but it is realized in the symmetric/antisymmetric-basis of the layer index. Consequently, the edge spectrum at a fixed particle number reveals an unexpected U(1)$_4 \times$ U(1) structure.
\end{abstract}

\maketitle

\section{Introduction}

The emergence of topological order is one of the most intriguing phenomena in interacting quantum systems.\cite{Wen90} Most importantly perhaps, emergent quasiparticles in two-dimensional topological phases of matter can acquire non-Abelian statistics and may provide quantum states with highly nonlocal entanglement that form an ideal basis for quantum information processing.\cite{TQCReview}   
Many unanswered questions about topological systems remain, despite recent developments in the field exploring phase transitions (notably those driven by topological Bose condensation)\cite{bais02,bais03,bais09a,bais09b,BarkeshliCondensation10, BurnellCondensation, barkeshli11-orbifold, Burnell12}, stability of topological phases to perturbations,\cite{Kitaev03, Hastings05, Hamma09, KlichAnnals10, Dusuel11} coupling of multiple non-Abelian subsystems,\cite{perspectives, EisensteinReview14, Vaezi14, mansson13}
or creation of non-Abelian theories from coupling simpler subsystems.\cite{Kane02, Teo14, Kraus13, Buehler14, Neupert14}
It is in these general realms that the current paper seeks to explore.

In most cases we are not able to easily relate the complex topological physics to more traditional condensed matter systems.  However, one important exception is that topological systems of Ising type (meaning they are described by a topological field theory related to the Ising Conformal Field Theory (CFT) or the SU(2)$_2$ Chern-Simons theory) can be frequently related to superconductors, thus providing a particularly powerful handle for understanding them. Such systems are now of  particular interest due to a variety of recent experiments aimed at realizing them in the laboratory.\cite{Dolev08,Radu08,Willett09,Kang11,WillettReview, LutchynMajoranas, AliceaMajoranas, Kouwenhoven12, Heiblum12, Deng12}  
In the current work we will examine a variant of the Moore-Read\cite{moore91} quantum Hall state, which is of this superconducting type.\cite{Greiter1991,ReadGreen00}

Another approach that has been extremely important in developing an understanding of topological phases is the use of exactly solvable models.\cite{Laughlin83, Haldane83, RezayiRead88, read96, ReadRezayi99, Davenport2012,  Kitaev03, Kitaev_Honeycomb, LevinWen05, WalkerWang, Keyserlingk13}  Even when exactly solvable models are very far from any real experimental system, their solutions teach us general principles, and we may hope that the physical systems will be described by the same phase of matter as the model, and will therefore have the same universal properties.   Further, with modern quantum technologies, such as cold atoms, trapped ions, or Josephson junctions,\cite{Zhang08, CooperShlyapnikov09, Han09, Sato09, Blatt08_Ions, Monroe13_Ions, JosephsonJunctionReview, Clarke_SCQubits08} one may hope that the precise model system may even be successfully realized in the future.  In this spirit, we will deploy model Hamiltonians that can be solved exactly as a central part of our current work.

In this paper, we consider the effect of inter-layer tunneling on a bilayer quantum Hall system formed by two bosonic $\nu=1$ Moore-Read states, each one being the exact ground state of a three-body contact interaction. While we frame the discussion in terms of a quantum Hall bilayer, similar considerations apply to any system with two internal degrees of freedom, including valley degrees or spin degrees of freedom.  To a large extent the same physics will occur also for interacting particles in Chern bands with Chern number $C=2$. 

Since the Moore-Read state can be thought of as a (chiral) $p$-wave superconductor of composite fermions, for intuition, it is useful to think of the inter-layer tunneling as a process occurring between two superconductors.   As is well known, tunneling of single particles is suppressed due to the superconducting gap, and one must consider then the tunneling of pairs, which gives rise to the rich  phenomenology of the Josephson effect.\cite{JosephsonPaper, JosephsonBook} In the case of coupled Moore-Read states, however, it is crucial that the paired particles are {\it composite fermions} --- in this case, bosons bound to (Jastrow factor) correlation holes.   Due to these correlation holes, it would be very difficult for bare bosons (paired or otherwise) to tunnel between the layers, as one must open a (Jastrow) correlation hole in the new layer and remove the (Jastrow) correlation hole from the old layer --- essentially moving the flux between the layers along with the boson.    

In order to create a setting for Moore-Read states in which tunneling is possible, we introduce an inter-layer correlation hole $\prod_{i,j} (z_i - w_j)$ by adding a suitably strong two-body inter-layer contact repulsion ($V_0$ of Haldane pseudo potentials\cite{Haldane83}).  In such a situation, a correlation hole always exists in the opposite layer, which can receive a tunneling particle easily. While one still expects single-particle tunneling to be suppressed due to the pairing physics of the Moore-Read state, in this situation one expects to realize pair tunneling similar to that of the conventional Josephson effect.    The exact ground state of our three-body intra-layer contact interaction along with the two-body inter-layer interaction is the coupled Moore-Read state, first discussed in Ref.~\onlinecite{prl12} In the resulting model, all particles carry the same Jastrow correlations. Hence, by removing the overall flux attachment, the system yields a solvable model for two Josephson coupled $p$-wave superconductors.

In the presence of pure three-body contact interactions plus inter-layer two-body repulsion, the coupled Moore-Read states are degenerate with respect to moving pairs of bosons between layers. This symmetry 
gives rise to a
Goldstone mode in the spectrum. Here, pair tunneling is crucial and even at infinitesimal magnitude it selects one particular ground state from the previously degenerate manifold of ground states and gaps the Goldstone mode. 
If we denote the pseudospins of bosons in the two layers with $\uparrow$ and $\downarrow$, then in the basis of symmetric and antisymmetric pseudospin states, $|\pm\rangle \propto |\uparrow\rangle \pm |\downarrow\rangle$, we find that the ground state wave function for small tunneling yields a particular superposition of coupled Moore-Read states that is exactly the Halperin 220 state.\cite{halperin83} This is a surprising result, for it immediately follows that we can write a purely two-body Hamiltonian which reproduces the exact ground state of our three-body interaction. 

Beyond the ground state properties, we see that the quasihole spectrum of these Hamiltonians can be calculated exactly and we demonstrate that it reflects the physics of confinement of quasiholes. To make this connection, we begin by calculating the full spectrum (for any number of quasiholes) of the system of uncoupled Moore-Read states, where the quasihole excitations can be viewed as vortices of the composite fermion superconductor.   In addition, one needs to consider the possible fusion channels of the Majorana zero modes in the vortex cores.\cite{moore91}  The introduction of Josephson tunneling between the layers can then be understood as locking the superconducting phase between the two layers, which results in binding (confining) vortices into pairs between the two layers, i.e.~each vortex is accompanied by a partner at the same location in the other layer. Taking this restriction into account, we find the spectrum of the coupled Moore-Read layers: the bulk is described by a U(1)$_2\times$U(1)$_2$ CFT, but the edge revels a different U(1)$_4\times$U(1) structure at fixed particle number. We show how these theories relate to the Ising$\times$Ising %field 
content of two uncoupled Moore-Read states, and demonstrate that the confinement of quasihole operators renders several types of topological excitations topologically invisible. The overall effect of the Josephson coupling on the spectrum of topological  bulk excitations is then the same as the effect that would be induced by topological Bose condensation of these invisible particles.

This paper is structured as follows: In section \ref{sec:JosephsonCoupling} we introduce our specific model of Josephson coupled Moore-Read states and discuss its qualitative physics and the nature of its ground state sector. In section \ref{sec:excitations}, we analyze the finite size quasihole spectra of our model bilayer Hamiltonians with Josephson coupling and also deduce the edge spectrum for an infinite droplet. In section \ref{sec:cft} we develop a description of this physics in terms of topological Bose condensation transitions between the underlying conformal field theories. The final section \ref{sec:conclusions} is devoted to conclusions and a general outlook for generalizations of our approach.

\section{Josephson Coupling for Paired Hall States}
\label{sec:JosephsonCoupling}

Our aim is to study the effect of Josephson coupling on a bilayer system composed of two Moore-Read states. Throughout this work, we focus on the case of bosons, although many of our results can be generalized to the case of fermions. Within each layer, we can use three-body contact interactions as the parent Hamiltonian of the Moore-Read state to induce the desired pairing properties. In order to create a ground state that is susceptible to tunneling, we additionally require a correlation hole between the layers. These correlations are created by an additional inter-layer repulsion in the Hamiltonian, leading us to define the parent Hamiltonian for coupled Moore-Read states, defined on the plane as
\be 
\label{eq:projectiveH}
\hat {\mathcal H}_{3\text{-}2} &=& \lambda_3 \sum_{i<j<k = 1}^{N_\uparrow} \delta^{(2)}(z^\uparrow_i - z^\uparrow_j)\delta^{(2)} (z^\uparrow_j - z^\uparrow_k) \nn\\&+&  \lambda_3\sum_{i<j<k = 1}^{N_\downarrow} \delta^{(2)}(z^\downarrow_i - z^\downarrow_j)\delta^{(2)} (z^\downarrow_j - z^\downarrow_k) \\\nn&+& \lambda_2 \sum_{i=1}^{N_\uparrow} \sum_{j=1}^{N_\downarrow} \delta^{(2)}(z^\uparrow_i - z^\downarrow_j).
\ee
We choose the prefactors $\lambda_2$ and $\lambda_3$ such that term acts as a projector on lowest Landau level states, and we use an equivalent construction in the spherical geometry for our numerics (see Appendix \ref{apx:sphere}).  The Hamiltonian (\ref{eq:projectiveH}) conserves the number of particles per layer, and its ground state wave function in a given sector with (even) number of particles $N_\uparrow$ and $N_\downarrow$ per layer is given -- as per construction -- by the coupled Moore-Read state that was introduced in Ref.~\onlinecite{prl12},
\be 
\Psi_0^{N_\uparrow, N_\downarrow}(\{z^\uparrow_i\},\{z^\downarrow_j\}) = {\rm Pf}\left(\frac{1}{z^\uparrow_i-z^\uparrow_j}\right){\rm Pf}\left(\frac{1}{z^\downarrow_i-z^\downarrow_j}\right)\\\nn
\times \prod_{i<j=1}^{N_\uparrow}(z^\uparrow_i - z^\uparrow_j)\prod_{i<j=1}^{N_\downarrow}(z^\downarrow_i - z^\downarrow_j)\prod_{i=1}^{N_\uparrow}\prod_{j=1}^{N_\downarrow}(z^\uparrow_i - z^\downarrow_j).
\label{eq:doublePfaffian}
\ee
This state has an overall filling factor of $\nu = \nu_\uparrow + \nu_\downarrow = 1$, and a shift of $-2$, i.e.~it includes orbitals up to a maximum flux of $N_\phi = N-2$.
We shall take polynomials in complex coordinates $z_j = x_j+i y_j$ to either denote states on the plane (omitting overall gaussian factors) or on the sphere (via stereographic projection), as explained further in Appendix \ref{apx:sphere}.

Each of the two coupled Moore-Read states can be thought of as a $p$-wave superconductor of composite fermions,\cite{moore91,ReadGreen00,MollerSimon08}
and we can deploy the picture of a Bose condensate of Cooper pairs. An important consequence is the fact that an odd number of particles per layer cannot be accommodated in the ground state. Instead, odd number configurations leave an unpaired fermion or `broken pair' which implies a finite gap $\Delta_\Psi$ for the resulting Bogoliubov quasiparticle or neutral fermion excitation.\cite{Greiter1991,Greiter1992,Bonderson11,Moller11} Hence, Hamiltonian (\ref{eq:projectiveH}) constrains the ground state to an even number of particles per layer. 

All sectors with an even number of bosons per layer possess an exact (and unique) zero energy ground state. It follows that (Cooper) pairs can be moved between the layers at vanishing energetic cost and, if we think in terms of the larger Hilbert space allowing such processes, the ground state degeneracy amounts to $d_{3\text{-}2} = N/2+1$.    Henceforth, let us denote the ground state with $N_\uparrow= 2n$ particles in the upper layer and $N_\downarrow=N-2n$ in the lower layer via the shorthand
\be
\Psi_{0,n} = \Psi_0^{2n,N-2n},\quad n = 0,\ldots,N/2.
\ee
As a consequence of the extensive ground state degeneracy, we need to include tunneling of (pairs of) particles between the layers in order to obtain the full physical picture. 
Taking the analogy with the superconducting system further, the number of Cooper pairs per layer is the conjugate variable to the phases $\chi_\sigma(\mathbf{r})$ of the superconducting order parameters. There is a U(1) symmetry in the difference $\Delta \chi = \chi_\uparrow - \chi_\downarrow$ between these complex superconducting order parameters, and we expect that long wavelength fluctuations of $\Delta\chi$ give rise to a Goldstone mode. Indeed, our numerics confirm that the gap of low-lying excited states at small angular momentum scales as $\Delta_\text{coll} \sim 1/N$, indicating the presence of a linearly dispersing Goldstone mode. Due to the discrete number of available momenta in finite size systems, the Goldstone excitations always occur at finite energy. Hence, we will focus on the physics of the ground state and zero-energy excitations for the coupled Moore-Read state, and leave the full exploration of the superconducting coherence and the collective Goldstone mode for a future publication.

To explore the effect of Josephson coupling, it would be sufficient to add a single-particle tunneling term $s (a^\dagger_\uparrow a_\downarrow + h.c.)$ to the Hamiltonian. While this term cannot move single particles between the layers due to the neutral fermion gap $\Delta_\Psi$, it would induce pair tunneling of magnitude $t\sim s^2 / \Delta_\Psi$ at the second order of perturbation theory. For simplicity, we extend our model directly by a pair tunneling term, and choose this to be local, i.e."we consider tunneling of pairs with relative angular momentum zero, given by the term
\be
\label{eq:VUmklapp}
\hat V_0^\text{tun}  =  \int d^2r  \hat \Pi_0^{\dagger}(\uparrow\uparrow,\mathbf{r}) \hat \Pi_{0}(\downarrow\downarrow, \mathbf{r}) + h.c.,
\ee
where $\hat \Pi_m^{\dagger}(\sigma\sigma',\mathbf{r})$ creates a pair of particles with individual spins $\sigma$, $\sigma'$, and relative angular momentum $m$ at the centre of mass position $\mathbf{r}$ [and $\hat \Pi_{m} = (\hat \Pi_{m}^\dagger)^\dagger$]. Explicit forms on the sphere are also given in Appendix \ref{apx:sphere}. Specifically, we consider the class of `Josephson coupled' Hamiltonians
\be
\label{eq:HamTot}
\hat {\mathcal H}_\text{3-2}^\text{JC} (t) = \hat {\mathcal H}_\text{3-2} + t \hat V_0^\text{tun},
\ee
The ground state of $\hat {\mathcal H}_\text{3-2}^\text{JC} (t=0)\equiv \hat {\mathcal H}_\text{3-2}$ is exactly degenerate and has a finite quasiparticle gap (and a finite collective mode gap $\Delta_\text{coll}\sim \mathcal{O}(1/N) $ in finite size systems of $N$ particles). We find that for small enough $t$, the tunneling term mixes only the $d_{3\text{-}2}$ zero-energy ground states, $\Psi_{0,n}$, so the new ground state can be obtained by degenerate perturbation theory. 

Let us think about tunneling using the picture of a superconducting system. As long as the tunneling strength vanishes, each of the degenerate eigenstates in the ground state sector $\Psi_{0,n}$ carries an arbitrary phase of $e^{i\phi_n}$ that we can understand as the finite-size equivalent of the order parameters $\chi$. As tunneling is turned on, the specific phase relationship with relative phases $\phi_n-\phi_{n-1} = \pi$ are selected, as these minimize the tunneling energy.\cite{NoteTunnelingSign} Let us assume for the moment that the superposition has equal weight for the different particle number sectors; we will verify this assumption numerically, below. In this case, the resulting state can be written as a triplet paired state with $\mathbf{d}$-vector $\propto \mathbf{e}_x$ (for details, see Appendix \ref{apx:mapping220}) as
\begin{align}
\label{eq:CoupledPfaffianDVector}
\nn
\Psi_0(t>0)%|_{(\{z_i^{\uparrow}\},\{z_i^{\downarrow}\})} 
= ~2^{-\frac{N}{2}}  {\rm Pf}\left[  \frac{\state{\uparrow\uparrow} - \state{\downarrow\downarrow}}{z_i - z_j} \right] \prod_{i,j} (z_i - z_j)\\
 =~ \sum_{n = 0}^N (-1)^n\Psi_{0,n},~~~~~~~~~~~~~~~~~~~~~~~~
\end{align}
to linear order in degenerate perturbation theory, i.e., $t\ll \Delta_\Psi$. 
Here, $z_i$ can stand for the position of a particle of either spin, which is assigned by the corresponding term in the pair wave function, in analogy to the work on fermionic bilayer systems by Ho.\cite{Ho95} Further using Ho's results,\cite{Ho95} we can show that (\ref{eq:CoupledPfaffianDVector}) is identical to the Halperin 220 wave function under a pseudo-spin rotation into the basis
\be
\label{eq:BasisChange}
\state{\pm} = \frac{1}{\sqrt{2}}\left( \state{\uparrow} \pm \state{\downarrow} \right).
\ee
With this change of basis, the pair wave function can be rewritten as $\state{\uparrow\uparrow} - \state{\downarrow\downarrow} = \state{+-} + \state{-+}$. Hence, we have 
\begin{align}
\Psi_0(t>0)%|_{(\{z_i^{+}\},\{z_i^{-}\})} 
=& ~2^{-\frac{N}{2}}  {\rm Pf}\left[  \frac{\state{+-} + \state{-+}}{z_i - z_j} \right] \prod_{i,j} (z_i - z_j) \nonumber\\
	=& ~{\rm Pf}\left[  \frac{1}{z_i^{+} - z_j^{-}} \right] \prod_{i,j} (z_i^+ - z_j^-)  \nonumber\\
	   & \times \prod_{i<j} (z_i^+ - z_j^+) \prod_{i<j} (z_i^- - z_j^-),
\end{align}
and by virtue of the Cauchy identity, 
\begin{align}
&{\rm Pf} \left[\frac{1}{(z_i^{+} - z_j^{-})}\right] \prod_{i,j} (z_i^{+} - z_j^{-})\\
=&\prod_{i<j}(z_i^{+} - z_j^{+})\prod_{i<j}(z_i^{-} - z_j^{-}),\nn
\end{align}
we confirm that 
\be
\Psi_0(t>0)|_{(\{z_i^{+}\},\{z_i^{-}\})} \equiv \Psi_{220}(\{z_i^{+}\},\{z_i^{-}\}).
\ee
Now, we know that the parent Hamiltonian for the $\Psi_{220}$ Halperin state is given by a contact interaction in each layer,
\be
\label{eq:Ham220}
\hat {\mathcal H}_{220}^{+-} = \hat V_0^{++} + \hat V_0^{--},
\ee
written in terms of the zeroth pseudopotential operators\cite{Haldane83} (see Appendix \ref{apx:sphere}). By applying the inverse basis change (\ref{eq:BasisChange}), we infer that there exists a parent Hamiltonian for the particular superposition, $\Psi_0(t>0)$, of the coupled Moore-Read states in the $\uparrow\downarrow$ basis, which is given by
\be
\label{eq:Ham220UD}
\hat {\mathcal H}_{220}^{\uparrow\downarrow} = \hat V_0^{\uparrow\uparrow} + \hat V_0^{\downarrow\downarrow} + \hat V_0^{\uparrow\downarrow} + \hat V_0^\text{tun}.
\ee
Again, we take units such that the $\hat V_0^{\sigma\sigma}$ term is a projector, as discussed in Appendix \ref{apx:sphere}. For an explicit derivation of the basis transformation leading to (\ref{eq:Ham220UD}), and for the definition of the inter-layer contact interaction $\hat V_0^{\uparrow\downarrow}$, see Appendix \ref{apx:mapping220}.  

\begin{figure}[t]
\begin{center}
\includegraphics[width=0.98\columnwidth]{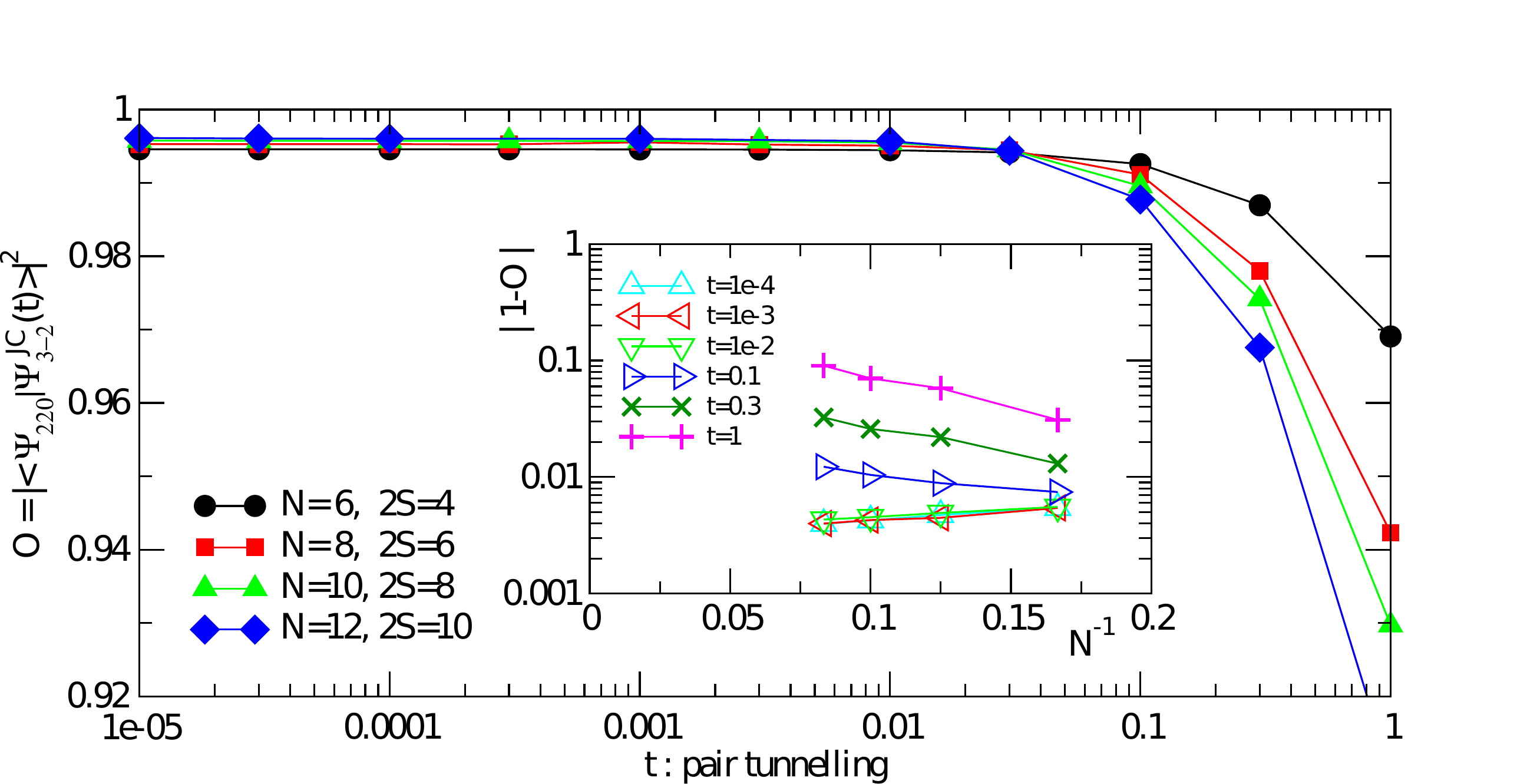} %{Overlaps_220_H32-t.pdf}
\caption{Overlaps $\mathcal{O} = |\langle \Psi_{220}^{\uparrow\downarrow} |�\Psi_\text{3-2}^\text{JC}(t)\rangle|^2$ between ground states of the three-body Hamiltonian with Josephson coupling (\ref{eq:HamTot}) and the 220 state in the $\uparrow\downarrow$ basis, the ground state of (\ref{eq:Ham220UD}). Data is shown for systems with $N=6, \ldots, 12$ particles on the Haldane sphere. Inset: scaling of the difference $|1-\mathcal{O}|$ with system size, highlighting the change in behaviour: the overlap decreases with $N$ in systems with $t\gtrsim 0.03$, while it increases for $t\lesssim 0.03$. }
\label{fig:OverlapsJC-220}
\end{center}
\end{figure}

We are now in the position to verify the claim that the 220 state is also generated by $\hat {\mathcal H}_\text{3-2}^\text{JC} (t\to 0)$ by comparing the ground state wave functions of (\ref{eq:HamTot}) and (\ref{eq:Ham220UD}). For system sizes up to $N=10$, and $t=10^{-4}$ we have verified that the ground state of $\hat {\mathcal H}_\text{3-2}^\text{JC} (t)$ are indeed superpositions involving only the zero energy ground states of $\hat {\mathcal H}_\text{3-2}$ to within a precision of $10^{-13}$. 
Fig.~\ref{fig:OverlapsJC-220} shows the overlaps of the exact ground states with the $220$ state for finite size systems on the sphere as a function of the tunneling parameter $t$.  For $t\gtrsim 0.03$, we find that the overlap is decreasing with system size, and for large $t$ there is some admixture of states beyond degenerate perturbation theory. However, for values of tunneling $t\lesssim 0.03$, we find that the overlap with 220 \emph{increases} with $N$.
On the basis of this data and our earlier heuristic arguments, we conjecture that the ground state of the three-body Hamiltonian with Josephson coupling tends exactly to the 220 state in the limit where $t$ goes to zero and $N$ is taken to infinity. 

Our considerations of pair tunneling terms in a two-body Hamiltonian were originally inspired by the possibility of having umklapp scattering in the Hofstadter lattice in Ref.~\onlinecite{prl12}. Indeed, the Hofstadter lattice provides a useful platform for strongly correlated quantum Hall liquids,\cite{KolRead, palmer06, PalmerJakschPRA, Moller09} leading up to the general formulation of fractional insulators in general Chern bands.\cite{Haldane88, ShengComm, Mudry11, RegnaultPRX, Scaffidi12, Sun12, Bergholtz13}
For Chern number $C=2$ bands, it is convenient to consider the single-particle eigenstates in terms of multiple flavors,\cite{RegnaultFlavours} and near $n_\phi=1/2$ the Hofstadter model provides bands with $C=2$ matching the two flavors of our current model.\cite{prl12}
However, a recent study\cite{harper14} argues that the microscopic Hamiltonian in the Hofstadter model does not yield a regime with sufficiently strong umklapp terms that could be described by two-body Hamiltonians of the type (\ref{eq:Ham220UD}), as originally believed.
Partially motivated by these (approximate) physical realizations, and partially in order to show that the solvable point of the 220 state belongs to a a wider range of parameter space, hence representative of a stable phase of matter, we also consider a generalized Hamiltonian where we allow for the magnitude of the tunneling term to be tuned by a variable prefactor $\alpha$:\cite{NoteHEff}
\be
\label{eq:HamEffAlpha}
\hat {\mathcal H}^\text{eff}(\alpha) = \hat V_0^{\uparrow\uparrow} + \hat V_0^{\downarrow\downarrow} + \hat V_0^{\uparrow\downarrow} +  \alpha \hat V_0^\text{tun}.
\ee
We discuss the excitation spectrum of this Hamiltonian in section \ref{sec:HamEffAlpha}.

Unlike $\hat {\mathcal H}_\text{3-2}^\text{JC}(t)$, which produces a gap of order $t$, the parent Hamiltonian $\hat {\mathcal H}_{220}^{\uparrow\downarrow}$, and its generalization $\hat {\mathcal H}^\text{eff}(\alpha)$, yield a large many-body gap and do not have a low-lying collective mode. The first two of these Hamiltonians share a common spectrum for the ground state and quasihole excitations, which can be calculated analytically. Analyzing the ground state sectors of the pure three-body Hamiltonian $\hat {\mathcal H}_{3\text{-}2}$ in comparison to the case of finite tunneling will inform us about the details of the transition that occurs when tuning from $t=0$ to finite tunneling $t$. The physical picture that emerges is the following: the quasiholes of the Moore-Read state can be thought of as vortices of the underlying composite fermion superconductor, i.e., the superconducting phase winds by $2\pi$ when going around one of these quasiparticles. In the absence of tunneling, we are free to place vortices independently in the $\uparrow$ and $\downarrow$ layers. However, when $t\neq 0$ we induce an 
energetic confinement of the $\uparrow$-vortices at $w_i^\uparrow$ to $\downarrow$-vortices at $w_j^\downarrow$, because an isolated vortex would create a mismatch of the superconducting phases between the two layers. Indeed, our analysis of the quasiparticle states confirms that the exact zero energy quasihole states of $\hat {\mathcal H}_{220}^{\uparrow\downarrow}$ are obtained from the quasihole spectrum of $\hat {\mathcal H}_{3\text{-}2}$ by identifying the quasihole positions $w_i^\uparrow= w_i^\downarrow$ (see Appendix~\ref{apx:exact} for an explicit example).

\section{Excitations}
\label{sec:excitations}
The formation of quasiparticle excitations is a hallmark of topological order in incompressible fractional quantum Hall liquids. As the (charge-) density of the system is perturbed from the preferred value, which is realized in the many-body ground state, the excess density gives rise to local deviations from the average that behave as emergent types of particles with fractional charge and fractional statistics. In the bosonic paired Hall states, described by Moore-Read's wave function at filling fraction $\nu = 1$, the quasiparticle charge amounts to $q_\text{qp}=\pm e/2$. In finite size systems, quasiparticle excitations can be studied either by changing the total number of particles in the system or, as we shall proceed, by adding/removing flux. This procedure changes the density by varying the overall area of the system while keeping the number of particles constant.

We focus on quasihole excitations that are obtained when additional flux is added. The specific number of low energy states at each angular momentum can be used as a probe of the underlying topological order.\cite{WenEdge90, MilovanovicRead} In the limit of a large droplet, the high angular momentum states yield a universal counting of edge states for the system on a disk, which can be derived from a $1+1$ dimensional Conformal Field Theory (CFT) describing the edge physics.\cite{WenEdge90}

Note that here we have two Hamiltonians, Eqs.~(\ref{eq:projectiveH}) and (\ref{eq:Ham220UD}), which have very similar ground states, in the sense that the projections of the ground states on fixed particle numbers in the layers are equal, but as we shall see, they give rise to different edge spectra. Nevertheless we argue that this observation does not pose a contradiction to the bulk-edge correspondence for topological systems, which states that in such systems, the edge and quasiparticle excitations follow directly from the ground state. Despite their similarity, our two Hamiltonians do not in fact have the same space of ground states: one has many degenerate ground states while the other does not. Furthermore, the Hamiltonian without tunneling, Eq.~(\ref{eq:projectiveH}), has gapless Goldstone modes and hence does not represent a proper topological phase of matter. The situation encountered here is similar to that which occurs in loop models based on $d$-isotopy\cite{Freedman08}, where the same ground state can be found both in a gapless phase and in a gapped topological phase.

\subsection{Quasihole spectrum of the effective projective Hamiltonian $\hat {\mathcal H}_{3\text{-}2}$}
\label{sec:h3-2}
As our target state (\ref{eq:doublePfaffian}) is the exact ground state of the projective three-body Hamiltonian (\ref{eq:projectiveH}), our aim is first to count the number of zero-energy quasihole excitations that this Hamiltonian supports at a given flux $N_\phi$. Their number follows directly from the counting of quasihole states in the single-component Moore-Read state of bosons at $\nu=1$, as first described in Ref.~\onlinecite{read96}. We briefly review these arguments in Appendix~\ref{apx:mr} and focus on the excitations of our two-component wave function (\ref{eq:doublePfaffian}) here.

In the case of the two-component system, described by Hamiltonian (\ref{eq:projectiveH}), we can determine the counting of quasiholes by directly using the results of Ref.~\onlinecite{read96}. The inter-layer correlations are screened perfectly already in the ground state (\ref{eq:doublePfaffian}), and this is not altered by adding quasihole excitations. This implies that there is no additional correlation between the quasihole positions in the two layers, and the wave function has the form
\be
\label{eq:MRxMRstates}
 \Psi^\text{qh}_{3\text{-}2} (\{z_i^\uparrow\}, \{z_j^\downarrow\}; \{w_k^\uparrow\}, \{w_l^\downarrow\}) ~~~~~~~~~~~~~~~~~~~~~~~~~\nn\\
 = \Psi^{\text{qh}, \:\nu = 1}_{m_1^\uparrow,\ldots m_{N_\uparrow}^\uparrow} (z_1^\uparrow,\ldots z_{N_\uparrow}^\uparrow; w_1^\uparrow,\ldots w_{2n_\uparrow}^\uparrow) \nn\\
 \times ~\Psi^{\text{qh}, \:\nu = 1}_{m_1^\downarrow,\ldots m_{N_\downarrow}^\downarrow} (z_1^\downarrow,\ldots z_{N_\downarrow}^\downarrow; w_1^\downarrow,\ldots w_{2n_\downarrow}^\downarrow) \nn\\
 \times \prod_{i,j} (z_i^\uparrow-z_j^\downarrow). ~~~~~~~~~~~~~~~~~~~~~~~~~~~~~~
\ee
Here $\Psi^{\text{qh}, \:\nu = 1}_{\{m_i^s\}}$ are quasihole wave functions of a single-layer Moore-Read state at filling fraction $\nu = 1$ and $\{m_i^s\}$ are fermionic occupation numbers [see Eq.~(\ref{eq:ReadRezayiQH}) in Appendix~\ref{apx:mr} for details]. The total number of excited states ensues simply by convoluting the countings of two independent single-layer Moore-Read states, 
\be
\label{eq:deg_3-2}
d_{3\text{-}2}(N_\uparrow,N_\downarrow,n_\uparrow, n_\downarrow, L_z)~~~~~~~~~~~~~~~~~~~~~~~~~~~~ \nn\\
= 
\sum_{l = 0}^{L_z} 
d_\text{MR}(N_\uparrow,n_\uparrow,l) d_\text{MR}(N_\downarrow,n_\downarrow,L_z - l)~~~
\ee
where $d_\text{MR}(N, n, L_z)$ denotes the degeneracy of a single-layer Moore-Read state with $N$ particles and $n$ additional flux quanta above the ground state at fixed angular momentum $L_z$ [see Eq.~(\ref{eq:dMR})].

Similarly, the character of the CFT, describing the edge of the infinite system,\cite{MilovanovicRead} factorizes into the components,
\be
 \chi^{3\text{-}2}(N_\uparrow,N_\downarrow)   = \chi^\text{MR}(N_\uparrow) \times \chi^\text{MR}(N_\downarrow),
\ee
where $\chi^\text{MR}$ is the character of the CFT associated with each layer [see Eq.~(\ref{eq:SingleMRCharacter})]. For the sector with even parity in both components, the character reveals a rapidly growing count of excitations,
\be
\label{eq:MRxMRcharacter}
1 + 2 q + 7 q^2 + 16 q^3 + 39 q^4 + 82 q^5 + 173 q^6 + \mathcal{O}(q^7).~~~~~
\ee
In finite size systems, only the first few terms of this series are recovered, while the counting of high angular momentum states is reduced. To give some examples, Table~\ref{tab:DegMRxMR} provides the degeneracies of $L_z$ eigenmodes with respect to the maximum angular momentum ($\Delta m = L_z^\text{max} - L_z$) and their equivalent expression as angular momentum multiplets $\mathcal{L}$, as predicted by formula (\ref{eq:deg_3-2}). We give exemplary data for several system sizes with $N_\uparrow = N_\downarrow $ even.
\begin{table*}[ht]
\begin{center}
\begin{tabular}{ccccccclclc}
& $N$ && $N_s$ && $n$ && $\mathcal{L}_{3\text{-}2}$&& $ \{d_{3\text{-}2}(\Delta m)| \Delta m=0,1,\ldots\}$ &  \\
\hline 
& 8 && 4 && 1 && $0^{2} \oplus 1^{1} \oplus 2^{3} \oplus 3^{1} \oplus 4^{1}$ && {\bf 1, 2}, 5, 6, 8 & \\
\hline 
& 8 && 4 && 2 && $ 0^{6} \oplus 1^{5} \oplus 2^{13} \oplus 3^{9} \oplus 4^{11} \oplus 5^{5} \oplus 6^{5} \oplus 7^{1} \oplus 8^{1} $ && {\bf 1,  2, 7}, 12, 23, 32, 45, 50, 56&  \\
\hline 
& 20 && 10 && 3 && $ 0^{612} \oplus 1^{1686} \oplus 2^{2790} \oplus 3^{3646} \oplus 4^{4436} \oplus 5^{4918} \oplus 6^{5292} \oplus 7^{5356} \oplus \ldots$ && {\bf 1, 2, 7, 16}, 37, 74, 145, $\ldots, 65018$ & \\
\hline 
& 32 && 16 && 4 && $ 0^{280877} \oplus 1^{832424} \oplus 2^{1374860} \oplus 3^{1888416} \oplus 4^{2375092} \oplus 5^{2816815} \oplus \ldots$ && {\bf 1, 2, 7, 16, 39}, 80, 165, $\ldots, 93073659$& \\
\hline
\hline
\end{tabular}
\caption{Degeneracies of quasihole excitations for the bilayer Moore-Read type Hamiltonian $\hat{\mathcal H}_{3\text{-}2}$, Eq.~(\ref{eq:projectiveH}), at $\nu=1$ in finite size geometries such as the sphere or a finite droplet on the plane. Data derived from formula (\ref{eq:deg_3-2}) is shown for systems of $N$ particles, with $N_s=N_\uparrow = N_\downarrow$ and $n$ units of flux above the ground state, $N_\phi=N-2+n$. The column $\mathcal{L}_{3\text{-}2}$ indicates the structure of the zero-energy quasihole space in terms of angular momentum multiplets [following (\ref{eq:low_energy_hilbert})]. The last column gives the number of quasihole states $d_{3\text{-}2}$ in the $L_z$ basis and relative to the edge (see main text). The results illustrate how the finite size data converges against the infinite system results in Eq.~(\ref{eq:MRxMRcharacter}) with growing system size. Countings that are converged to this limit are highlighted in bold. The last entry under $d_{3\text{-}2}$ reflects the degeneracy in the $L_z=0$ sector, giving the total count of angular momentum multiplets at this system size.}
\label{tab:DegMRxMR}
\end{center}
\end{table*}

\begin{figure}[t]
\begin{center}
\includegraphics[width=0.8\columnwidth]{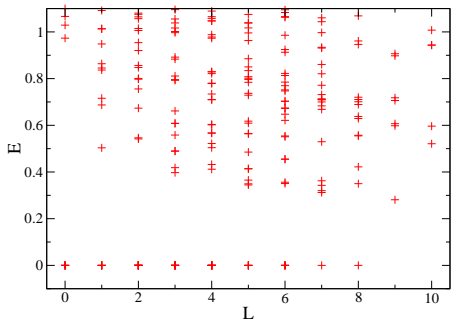}
\caption{Spectrum of the model Hamiltonian (\ref{eq:projectiveH}) for $N_\uparrow+N_\downarrow=4+4$ on a sphere with $N_\phi=8$ flux quanta, i.e., $n=2$ flux quanta above the ground state as a function of total angular momentum $L$. The counting of quasihole states in the zero-energy manifold matches the theoretical prediction given in Eq.~(\ref{eq:MRxMR_counting_N8_n_2}). We further identify the gap in this finite-size system to be $\Delta_{N=8}\simeq 0.2812$. }
\label{fig:SpectrumModelH23}
\end{center}
\end{figure} 

In addition to the ground state manifold of quasiholes, exact diagonalization yields the spectrum of excitations, so we can easily check our predictions with numerical tests. As an example, we consider the spectrum for the three-body Hamiltonian $\hat {\mathcal H}_{3\text{-}2}$ for $N=4+4$ particles with $n=2$ additional flux. For this system, there are 56 zero-energy states spanning the zero-energy Hilbert subspace $\mathcal{L}$ with a  structure of angular momentum multiplets ($L^2$ eigenstates) given by
\be
\label{eq:MRxMR_counting_N8_n_2}
\mathcal{L}=0^6 \oplus 1^5  \oplus 2^{13} \oplus 3^9 \oplus 4^{11} \oplus 5^{5} \oplus 6^{5} \oplus 7^{1} \oplus 8^{1},~~~
\ee
following the notation introduced in (\ref{eq:low_energy_hilbert}).
Indeed, we find that this is precisely the counting of states obtained in our exact diagonalization studies of the three-body model Hamiltonian (\ref{eq:projectiveH}) on the sphere, following the conventions that are well-established in the literature.\cite{Haldane83,MilovanovicRead} In order to numerically resolve the angular momentum structure of the degenerate zero-energy quasihole states, we diagonalize the angular momentum operator $L^2$ in the subspace spanned by eigenvectors of the corresponding degenerate eigenstates. 
Fig.~\ref{fig:SpectrumModelH23} shows the spectrum for the example considered above ($N=4+4$, $n=2$). We find that the zero-energy subspace of this finite size system is well separated from excited quasihole states by a gap: the lowest excited state occurs at $L=9$ and the gap is $\Delta_{N=8} \simeq 0.2812$. 
This finite size gap allows us to clearly define the zero-energy subspace. Note, however, that this Hamiltonian is gapless in the thermodynamic limit.

\subsection{Quasihole spectrum of the effective pseudospin Hamiltonian $\hat {\mathcal H}^{\uparrow\downarrow}_{220} $} 
\label{sec:QH_220}
We now turn to the excitation spectrum of Hamiltonian $\hat {\mathcal H}^{\uparrow\downarrow}_{220}$, whose exact ground state at $\nu=1$ is a particular equal-weight superposition of coupled Moore-Read states for all (fixed parity) particle number distributions between the two layers (Eq.~\ref{eq:CoupledPfaffianDVector}). Our numerical studies reveal that there are quasihole states that are exact zero energy ground states of this Hamiltonian, to within numerical accuracy. The existence of quasihole states with zero energy in the system is understood by the observation that an effective three-body repulsive behavior is generated by $\hat {\mathcal H}^{\uparrow\downarrow}_{220}$. Indeed, the quasihole wave functions of the Moore-Read state continue to satisfy the three-body vanishing property of the ground state, namely that the eigenstates vanish when three or more particles come together, while they can remain finite when only two particles come together. We explain this in more detail in Appendix~\ref{apx:exact}.

\begin{figure}[t]
\begin{center}
\includegraphics[width=0.8\columnwidth]{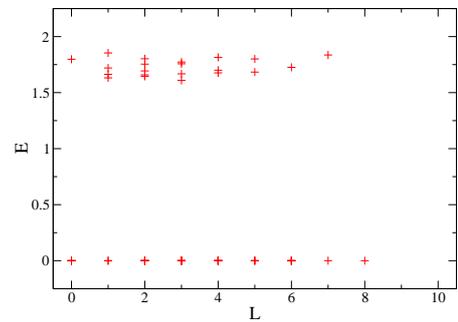}
\caption{Spectrum of the two-body contact interaction in the $\uparrow\downarrow$ basis, $\hat {\mathcal H}^{\uparrow\downarrow}_{220}$, for $N=8$ particles on a sphere with $N_\phi=8$ flux quanta, i.e., $n=2$ flux quanta above the ground state. The Hamiltonian has a large number of zero-energy states, but the degeneracy of this quasihole subspace is found to be lower than the corresponding degeneracy obtained for the case of $\hat {\mathcal H}_{3\text{-}2}$. The system features a large gap of $\Delta\simeq 1.609$ (known to be robust in the thermodynamic limit).}
\label{fig:SpectrumHEffAlpha1}
\end{center}
\end{figure} 

We now discuss the nature of the zero-energy eigenstates of $\hat {\mathcal H}^{\uparrow\downarrow}_{220}$ projected to sectors with fixed particle numbers per species, or fixed pseudospin $S_z = (N_\uparrow-N_\downarrow)/2$, denoting the corresponding projection operator by $\mathcal{P}_{S_z}$. This is analogous to our discussion of the ground state in section \ref{sec:JosephsonCoupling} and in Ref.~\onlinecite{prl12}. Having taken the $\mathcal{P}_{S_z}$ projection, we can compare the resulting wave functions to the eigenstates of the $S_z$-conserving  three-body model Hamiltonian (\ref{eq:projectiveH}). We find that the zero-energy eigenstates of $\hat {\mathcal H}^{\uparrow\downarrow}_{220}$ are fully contained within the basis of zero-energy eigenstates of $\hat {\mathcal H}_{3\text{-}2}$. Numerically, we find that for degenerate angular momentum multiplets, the quasihole states of the former are linear combinations of the quasihole states of the latter. Or, for non-degenerate angular momentum multiplets they are, in fact, identical. The remaining task is to identify which of the zero-energy eigenstates of $\hat{\mathcal H}_{3\text{-}2}$ are also zero-energy eigenstates of $\hat {\mathcal H}^{\uparrow\downarrow}_{220}$.

We have carried out numerical calculations for up to $N = 8$ particles and $n = 2$ flux quanta added. At this system size, we find $39$ zero energy eigenstates, of which $21$ have a non-zero projection onto the subspace with $N_\uparrow = N_\downarrow$, or pseudospin $S_z=0$. These states also have non-zero weight in the subspaces with $S_z \mod 2 = 0$, which are related to each other by pair tunneling. Note that the parity of the particle number is a symmetry of the Hamiltonian, so ($S_z \mod 2$) is conserved. This 21-fold degeneracy should be compared to the $56$ states arising for quasiholes of $\hat {\mathcal H}_{3\text{-}2}$ at this system size, as given in Table~\ref{tab:DegMRxMR}. Hence, the low lying band of the spectrum of $\hat {\mathcal H}^{\uparrow\downarrow}_{220}$ has \emph{fewer} low lying states than the quasihole excitation spectrum of two independent Moore-Read states resulting from Hamiltonian $\hat {\mathcal H}_{3\text{-}2}$. The remaining $18$ zero-energy eigenstates of $\hat {\mathcal H}^{\uparrow\downarrow}_{220}$ have support in the 
subspaces with odd $S_z$. Excited states are separated from the zero-energy manifold by a well-defined gap, as shown in Fig.~\ref{fig:SpectrumHEffAlpha1}.

The essential ingredient of $\hat {\mathcal H}^{\uparrow\downarrow}_{220}$ is the presence of the pair tunneling terms. It is instructive to think of the ground state wave function (\ref{eq:doublePfaffian}) as encoding the physics of two individual superfluids for particles with pseudospins up and down. But, in addition, these superfluids are mutually phase coherent, thanks to the Josephson coupling that arises from tunneling pairs in $\hat {\mathcal H}^{\uparrow\downarrow}_{220}$. In the presence of quasiholes, we claim that this phase coherence is crucial: when a quasihole is introduced into only one of the two pseudospin species, it causes a winding of the phase of the underlying superconducting order. This frustrates the Josephson coupling term and costs a large amount of energy. However, this cost can be eliminated entirely by introducing a quasihole that causes an equivalent phase winding of the order parameter for the other pseudospin species. From this picture, we should expect that it is favorable to introduce quasiholes at 
coinciding locations for both pseudospins. As this leaves fewer positional degrees of freedom for placing the quasiholes, the number of available zero-energy eigenstates is reduced.

To understand the implications of this picture on the excitation spectrum, let us look at the structure of the quasihole states within the framework that Read and Rezayi have established.\cite{read96} Similar to a single-layer Moore-Read state, the degeneracy of the quasihole states depends on two factors:  an orbital degeneracy, which is related to the number and positions of quasiholes, and a `topological degeneracy' (in the terminology of Read and Rezayi that we review in Appendix \ref{apx:mr}). The topological degeneracy can be expressed in terms of the number and orbital quantum numbers of unpaired fermions, or broken pairs, that characterize the quasihole sectors in the presence of additional flux.

In the case of the Josephson coupled bilayer Moore-Read state, the number and positions of quasiholes are constrained such that they must be the same in both layers, i.e.
\be 
\label{eq:constraintQH}
w_{i}^\uparrow = w_{i}^\downarrow \equiv w_i, \quad i=1,\ldots 2n.
\ee
Consequently, we deduce that the zero-energy eigenstates of (\ref{eq:Ham220UD}) must be of the form
\begin{align}
\label{eq:QHWaveFunctions}
\Psi_\text{JC}^\text{qh} (\{z_k^\uparrow\}, \{z_l^\downarrow\}; \{w_i\}) = \Psi^\text{qh}_{3\text{-}2} (\{z_k^\uparrow\}, \{z_l^\downarrow\}; \{w_i\}, \{w_i\}),
\end{align}
where the subscript JC may refer to both our `Josephson coupled'  Hamiltonians $\hat {\mathcal H}_\text{3-2}^\text{JC}(t\to 0)$ and $\hat {\mathcal H}_\text{220}^{\uparrow\downarrow}$,  as we expect that quasiholes are bound in both these systems. However, the quasihole wave functions (\ref{eq:QHWaveFunctions}) are exact zero energy eigenstates only for $\hat {\mathcal H}_\text{220}^{\uparrow\downarrow}$.
The constraint on quasihole positions (\ref{eq:constraintQH}) has no impact on either the number or the orbitals available to unpaired fermions in each of the Pfaffian states, so the expressions for the topological degeneracy can still be applied to each layer, individually. As per the arguments for the single-layer system in Appendix \ref{apx:mr}, we get independent factors for each of the pseudospin species, so topological degeneracies, $d_\text{topo}(n, p_\uparrow)$ and $d_\text{topo}(n, p_\downarrow)$, are given by Eq.~(\ref{eq:DTopo}), with $p_s$ the number of unpaired fermions for pseudospin~$s$. 

\begin{table*}[ht]
\begin{center}
\begin{tabular}{ccccclclclc}
& $N$  && n && $\mathcal{L}^\text{JC}_\text{even}$ && $\mathcal{L}^\text{JC}_\text{odd}$ && $\{d_\text{total}^\text{JC}(\Delta m)| \Delta m=0,1,\ldots\}$ &  \\
\hline 
& 6&& 1 && $0^1 \oplus 2^1$ && $1^1 \oplus 3^1$ && {\bf 1, 2}, 3, 4 & \\
\hline 
& 6 && 2 && $0^3 \oplus 2^4 \oplus 4^3 \oplus 6^1 $ && $ 0^1 \oplus 1^2 \oplus 2^2 \oplus 3^2 \oplus 4^2 \oplus 6^1 $ &&{\bf 1, 2, 7}, 10, 16, 18, 22 & \\
\hline 
& 6 && 3 && $1^6\oplus 2^4\oplus 3^9\oplus 4^5\oplus 5^6\oplus 6^3\!\oplus 7^3\!\oplus 9^1 $ && $0^3\oplus 1^3\oplus 2^7\oplus 3^6\oplus 4^7\oplus 5^4\!\oplus 6^4\oplus 7^2\!\oplus 8^1 $ && {\bf 1, 2, 7, 14}, 24, 36, 51, 62, 71, 74& \\
\hline 
& 8 && 1 && $0^{1} \oplus 2^{1} \oplus 4^{1}  $ && $ 1^{1} \oplus 3^{1} $ && {\bf 1, 2}, 3, 4, 5 & \\
\hline 
& 8 && 2 && $0^{4} \oplus 2^{5} \oplus 3^{2} \oplus 4^{5} \oplus 5^{1} \oplus 6^{3} \oplus 8^{1}$ && $0^{1} \oplus 1^{2} \oplus 2^{3} \oplus 3^{4} \oplus 4^{3} \oplus 5^{2} \oplus 6^{2} \oplus 7^{1}$ && {\bf 1, 2, 7}, 10, 18, 24, 32, 34, 39 &  \\
\hline 
\hline
\end{tabular}
\caption{Degeneracies of quasihole excitations for the effective Hamiltonian with Josephson pair-tunneling  $\hat {\mathcal H}^{\uparrow\downarrow}_{220}$,  Eq.~(\ref{eq:Ham220UD}), at $\nu=1$ in finite size geometries such as the sphere or a finite droplet on the plane. Data derived from (\ref{eq:degJC}) is shown for systems of $N$ particles at $n$ units of flux above the ground state, $N_\phi=N-2+n$. In analogy with table \ref{tab:DegMRxMR}, columns $\mathcal{L}^\text{JC}_\text{even/odd}$ give the number of angular momentum multiplets; additionally these are separated into sectors with an odd / even number of unpaired fermions. Finally, $\{d_\text{total}^\text{JC}(\Delta m)\}$ is the total number of states in the $L_z$ basis (of either parity) and with angular momentum relative to the edge.
The edge theory of the infinite system is described by the character (\ref{eq:JCCharacter}), and finite size data that are converged to this limit are highlighted in bold. The last entry under $d_\text{total}^\text{JC}$ reflects the degeneracy in the $L_z=0$ sector, giving the total count of angular momentum multiplets at this system size.}
\label{tab:DegEffectiveH}
\end{center}
\end{table*}

\begin{figure*}[t]
\begin{center}
\includegraphics[width=0.95\textwidth]{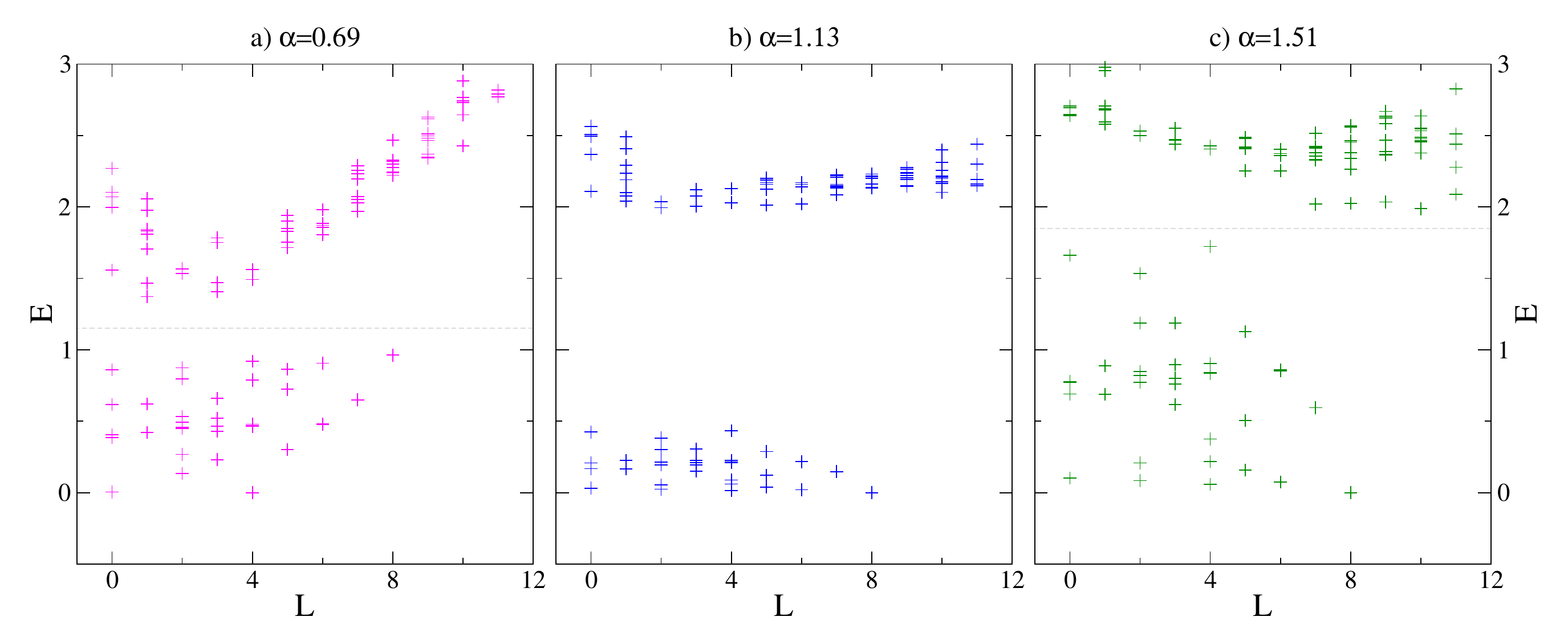}
\caption{Spectrum of the generalized pair tunneling Hamiltonian (\ref{eq:HamEffAlpha}) for $N=N_\uparrow+N_\downarrow=8$ on a sphere with $N_\phi=8$ flux quanta, i.e., $n=2$ flux quanta above the ground state as a function of total angular momentum $L$. Energies are given in natural units of $V_0$, relative to the lowest energy state at each $\alpha$. Spectra are shown for a) $\alpha=0.22\pi\simeq 0.69$, b) $\alpha=0.36\pi\simeq 1.13$, and c) $\alpha=0.48\pi\simeq 1.51$. For values $\alpha \simeq 1$, such as shown in panel b), there is a clear gap. Tuning further away from $\alpha=1$, the quasihole spectrum acquires more dispersion, and finally merges into the continuum of excited states. Panels a) and c) show spectra near the lower and upper boundary of the region in $\alpha$ for which the counting of quasihole states in the low-energy manifold still matches the count of states for $\hat {\mathcal H}_{220}$. Grey dashed lines highlight the separation of the low-energy manifold of states and other excited states.}
\label{fig:SpectrumHEffAlpha}
\end{center}
\end{figure*} 

Unlike the topological degeneracy, the constraint on the positions of quasiparticles in the two layers, Eq.~\ref{eq:constraintQH}, directly affects the orbital degeneracy, which follows from expanding the quasihole states in terms of symmetric polynomials of the $2n$ quasihole coordinates $\{w_i\}$ in each layer. For the single-layer case Read and Rezayi show that the relevant polynomials have a degree of at most $(N-p)/2$. In the present case we have to consider contributions from both layers, so the highest power of any one $w_i$ is
\be
r = \frac{N_\uparrow-p_\uparrow}{2} +  \frac{N_\downarrow-p_\downarrow}{2} = \frac{N-p_\uparrow-p_\downarrow}{2}.
\ee
The total number of homogeneous polynomials of the $2n$ quasihole coordinates $\{w_i\}$ with rank $r$ reduces to the known count of the degeneracy of $2n$ bosons filling a Landau level with $r$ flux quanta, which yields the orbital degeneracy of [cf. Eq.~(\ref{eq:DOrb}) in Appendix \ref{apx:mr}],
\be
\label{eq:HEffOrbitalDeg}
d_\text{orb}(N, n,p_\uparrow+p_\downarrow) = \binom{(N-p_\uparrow-p_\downarrow)/2 + 2n}{2n},
\ee

Using these results, the zero-energy quasihole space of Hamiltonian (\ref{eq:Ham220UD}) can be given by coupling the angular momenta pertaining to the two topological sectors and the common orbital sector as,
\be
\label{eq:degJC-0}
\nn D^\text{JC}(N,n) = 
\sum_{\substack{p_\uparrow, p_\downarrow\\
 \{N-p\: \text{mod}\: 2 = 0\}}}^{n} d_\text{topo}(n,p_\uparrow) \times d_\text{topo}(n,p_\downarrow) 
\nn\\
\times~ d_\text{orb}\left(N,n,p_\uparrow+p_\downarrow\right).~~~~~~~~
\ee
Noting that the topological (orbital) degeneracy is related to a filling of fermionic (bosonic) Landau-level orbitals, the spectrum can easily be evaluated as a function of total angular momentum $L_z$, and takes the form
\be
\label{eq:degJC}
\nn d^\text{JC}(N,n, L_z) = 
\sum_{\substack{p_\uparrow, p_\downarrow\\
 \{N-p\: \text{mod}\: 2 = 0\}}}^{n}
\sum_{l'=0}^{L_z} 
\sum_{l=0}^{l'}
 ~~~~~~~~~~~~~~~~~ \\
d_\text{Fermi}(p_\uparrow, n-1,l) \times d_\text{Fermi}(p_\downarrow, n-1, l' - l) ~~~~~~~~
\nn\\
\times~ d_\text{Bose}\left(2n, \frac{N-p_\uparrow-p_\downarrow}{2}, L_z - l'\right).~~~~~~~~~~~~~~~~~~~~
\ee
Here $d_\text{Bose/Fermi}(N,N_\text{orb}, L_z)$ refer to the degeneracies associated with placing $N$ Bosons/Fermions in $N_\text{orb}$ orbitals for states with fixed angular momentum $L_z$ [see Eqs.~(\ref{eq:dbose}) and~(\ref{eq:dfermi}) in Appendix~\ref{apx:mr}].

In the single-layer Moore-Read state, the number of unpaired fermions was constrained to values with the same parity as the number  of particles. In the Josephson coupled bilayer case, the bosonic factor in Eq.~(\ref{eq:degJC}) imposes only a single constraint, namely on the total number of particles $N$ and the total number of broken pairs $p_\uparrow + p_\downarrow$ to have the same parity. For even $N$, the two possible solutions consist of either having both $N_\uparrow$ and $N_\downarrow$  to be odd, or having both of them to be even. These two sectors remain uncoupled by the pair tunneling term (\ref{eq:VUmklapp}) in the Hamiltonian, so we can count each of these independently. Similarly, for odd $N$ the sum consists of two independent sectors where $N_\uparrow$ and $N_\downarrow$ have opposite parity.

We can now compare these results to exact diagonalization studies of the Hamiltonian $\hat {\mathcal H}^{\uparrow\downarrow}_{220}$. Table~\ref{tab:DegEffectiveH} lists some of the degeneracies associated with quasihole states at fixed angular momentum $L_z$ obtained in calculations carried out in spherical geometry. We obtain perfect agreement for the quasihole counting between numerical simulations and Eq.~(\ref{eq:degJC}) for all five system sizes shown in Table~\ref{tab:DegEffectiveH} and also in Fig.~\ref{fig:SpectrumHEffAlpha1}. For example, in the system of $N=8$ and $n=2$, we find that the $39$ quasihole states are distributed precisely according to the angular momenta and parity of the $S_z$ projection given in the table. All of the other systems shown are also perfectly matched. We conclude that the two-component Moore-Read system with pair tunneling is accurately described by our picture of Josephson coupling which enforces the identification of quasihole positions for the two spin components.

Let us now discuss the nature of the edge spectrum in the Josephson coupled Moore-Read state. The character describing the edge of a single Moore-Read Pfaffian state is determined entirely by the parity of the number of particles. The effective Hamiltonian with pair hopping conserves the parity in each layer. For a fixed total number of particles $N$ even, possible edge states include terms where the total number of particles per layer $N_\uparrow$ and $N_\downarrow$ have the same parity, either even or odd. Hence, the topological sector is described by the square of the Majorana-Weyl characters of even and odd parity. Simultaneously, the charge sector has a single Bose field just as in the single-layer case [see Eq.~(\ref{eq:mwchi})]. Correspondingly, the edge theory for even $N=N_\uparrow+N_\downarrow$ is described by the overall character,
\be
\label{eq:JCCharacter}
\chi^\text{JC} = \left[ (\chi^\text{MW}_+ )^2 + (\chi^\text{MW}_- )^2 \right] \chi^\text{Bose}
\ee
The first terms in this series are given by 
\be
1 + 2 x + 7 x^2 + 14 x^3 + 30 x^4 + 56 x^5 + \mathcal{O}(q^6),~~~~
\ee
and we see that the number of excitations is indeed growing more slowly than for the unconstrained case of Moore-Read $\times$ Moore-Read in (\ref{eq:MRxMRcharacter}).~\cite{footnote:oddN} The last column in Table~\ref{tab:DegEffectiveH} indicates the corresponding edge counting in finite size systems, indicating that the finite size results converge (slowly) towards the expected results. 

\subsection{Quasihole spectrum of the generalized Hamiltonian $\hat {\mathcal H}^\text{eff}(\alpha)$}
\label{sec:HamEffAlpha}

Beyond the solvable point $\hat {\mathcal H}^\text{eff}(\alpha=1)\equiv \hat {\mathcal H}_{220}^{\uparrow\downarrow}$ that was discussed in the preceding section \ref{sec:QH_220}, the phase which is described correctly by the physics of Josephson coupled quasiholes extends to a broader region in the parameter $\alpha$. We have previously shown overlaps for the ground state of the Hamiltonian (\ref{eq:HamEffAlpha}) with the coupled Moore-Read state in Fig.1c of Ref.~\onlinecite{prl12} (where $\alpha=2\pi\epsilon$, see note \onlinecite{NoteHEff}), revealing a broad maximum around the exactly solvable point $\alpha=1$. This behaviour suggests that the topological order remains robust over this entire region, and we confirm this explicitly by inspection of the quasihole spectra of the model. To illustrate this point, we present energy spectra for the quasiholes of $N=8$ particles and $N_\phi=8$, i.e., $n=2$ flux quanta above the ground state in Fig.~\ref{fig:SpectrumHEffAlpha}. Indeed, we find that the counting of quasihole states from table \ref{tab:DegEffectiveH} (last row), continues to apply throughout a range of about $0.7 \lesssim \alpha \lesssim 1.5$. In the centre of this region, the low-energy quasihole states are separated from higher excited states by a large gap.

The non-trivial dispersion of the quasihole spectrum away from the high symmetry point $\alpha=1$ signals that matrix elements of the pair-hopping terms depend on the specific quasihole wave function. Indeed, this is expected as the quasihole states have different numbers of broken pairs. Since we know that the  Hamiltonian $\hat {\mathcal H}_{220}^{\uparrow\downarrow}$ yields zero-energy ground states with Josephson coupled quasiholes, we can use this as a small perturbation to the three-body Hamiltonian $\hat {\mathcal H}_\text{3-2}$. Considering the linear combinations $\hat {\mathcal H}(\kappa) = \hat {\mathcal H}_\text{3-2} + \kappa \hat {\mathcal H}_{220}^{\uparrow\downarrow}$, one obtains a family of Hamiltonians that induces Josephson coupling between the layers \emph{without} distinguishing energetically between the various bound quasihole states satisfying (\ref{eq:constraintQH}).

\subsection{Intermediate Summary}

We have analyzed the excitation spectra of two different parent Hamiltonians, namely the three-body Hamiltonian $\hat {\mathcal H}_{3\text{-}2}$  and the two-body Hamiltonian with pair tunneling $\hat {\mathcal H}_{220}^{\uparrow\downarrow}$. The ground states of the former are coupled Moore-Read states with any even number of particles per layer, whereas the latter has a unique ground state that is an equal weight superposition of these coupled Moore-Read states with fixed particle number parity. Both Hamiltonians are exactly solvable in the sense that we can write down their exact ground state and zero-energy quasihole wave functions. 

We have verified numerically that the excitation spectrum of $\hat {\mathcal H}_{3\text{-}2}$ has the same number of zero energy states as two independent Moore-Read states in every angular momentum sector, while the spectrum of $\hat {\mathcal H}_{220}^{\uparrow\downarrow}$ grows at a slower rate. We have explained this smaller number of quasihole states in the picture of Josephson coupled superconductors, which enforces that quasiholes in one layer are bound to quasiholes in the other layer in order to minimize the Josephson energy. This picture is confirmed by a perfect match of the number of quasihole states, and allows us to explicitly construct their wave functions. 

We have further shown that this physics is robust to variations in the pair tunneling away from the exactly solvable models. For both cases, we have the universal edge counting in the thermodynamic limit that affords a classification of the topological order in each case. In the following section we analyze this classification from a CFT point of view. We will thus provide an interpretation for the edge spectrum and will demonstrate that Josephson coupling can be described in the CFT language by considering the confinement of individual quasihole fields.

\section{Conformal Field Theory description of the coupled Moore-Read state}
\label{sec:cft}

As was shown in Sec.~\ref{sec:h3-2} the counting of edge excitations of the three-body Hamiltonian $\hat {\mathcal H}_{3\text{-}2}$ matches the characters of a SU(2)$_2$ $\times$ SU(2)$_2$ Wess Zumino Witten CFT.\cite{bigyellowbook} The bulk wave function can also be produced from conformal blocks of this CFT, but only up to the inter-layer Jastrow factor. We have not been able to find a CFT that produces both the correct wave function, with this factor included, and the observed edge counting, and conjecture that such CFT may not exist. This observation is related to the fact that this Hamiltonian is inherently gapless and the finite gap observed in our numerical studies in the previous section is an artifact of using finite size systems. 

Conversely, the two-body Hamiltonian $\hat {\mathcal H}_{220}^{\uparrow\downarrow}$ is truly gapped at total filling fraction $\nu = 1$, hence in this case the coupled Moore-Read state represents a gapped topological phase of matter. Furthermore, as was shown in Ref.~\onlinecite{prl12} and section \ref{sec:HamEffAlpha} above, the ground state and excitation spectrum of this Hamiltonian are robust under variations of the tunneling strength away from unity, for the generalized two-body Hamiltonians with pair tunneling, Eq.~(\ref{eq:HamEffAlpha}). The aim of this section is to identify a CFT that describes both the edge excitation spectrum and the bulk wave function of this phase, in the spirit of Moore and Read.\cite{moore91}

In determining this CFT, and hence the topological order of this phase, we start from the non-Abelian Ising $\times$ Ising $\times$ U(1) picture, which is naturally suggested by the coupled Moore-Read wave function. In this picture, we will show that the lowest charge quasiparticles must consist of confined fields. Investigating the result of this confinement on the spectrum we then arrive at the equivalent CFT description in terms of a U(1) $\times$ U(1) CFT, which shows explicitly that the topological order of this phase is Abelian.

\subsection{The Coupled Pfaffian State and the Ising CFT}
Following the original work of Moore and Read, FQH wave functions can be expressed as correlators of the primary fields of certain conformal field theories.\cite{moore91} In this approach electron and quasihole operators are defined in terms of the primary fields of one or more CFTs. The ground state and quasihole wave functions are then calculated by evaluating the correlators of these operators.

In our case, the coupled Moore-Read state can be viewed as two single-layer Moore-Read Pfaffian states that share the same charge. Therefore, the natural starting point for the CFT is a combination of two Ising theories with primary fields denoted as $\{1,\sigma^\uparrow,\psi^\uparrow\}, \{1,\sigma^\downarrow,\psi^\downarrow\}$, to represent the neutral sector, and a chiral Bose field U(1)$_4$ [read: U(1) level 4] with primary fields $\{1, e^{i\phi_c/2}, e^{-i\phi_c/2}, e^{i\phi_c}\}$, to account for the charge sector of this state. Here the specified `level' $N = 4$, which represents the compactification of U(1)$_N$ with radius $R = \sqrt{N} = 2$, is related to the quantization of charge, with the smallest possible charge in the system being $e/2$ where $e$ is the unit charge of electrons. 

Similar to the CFT description of the Pfaffian state,\cite{moore91} we define bosonic `electron' operators for each layer,
\be 
\label{eops}
\nn
\psi_e^\uparrow(z^\uparrow) &=& \psi^\uparrow(z^\uparrow) e^{i\phi_c(z^\uparrow)},\\
\psi_e^\downarrow(z^\downarrow) &=& \psi^\downarrow(z^\downarrow) e^{i\phi_c(z^\downarrow)}.
\ee
The correlator of these fields reproduces the electronic ground state,\cite{footnote:bg charge}
\be
\Psi_0^{N_\uparrow,N_\downarrow}(\{z^\uparrow_i\},\{z^\downarrow_j\}) = \langle \prod_{i = 1}^{N_\uparrow}\prod_{j = 1}^{N_\downarrow}\psi_e^\uparrow(z_i^\uparrow) \psi_e^\downarrow(z_j^\downarrow)\rangle,
\ee
which gives the coupled Moore-Read state, Eq.~(\ref{eq:doublePfaffian}) for fixed numbers of particles in each layer. 

When considering the quasihole operators, we see that the smallest-charge quasihole operators that are local with respect to electrons can only be of the form,
\be
\psi_{qh}(w) = \sigma^\uparrow(w)\sigma^\downarrow(w) e^{i\phi_c(w)/2}.
\label{qhop}
\ee
This is similar to the quasihole operator of a single-layer Moore-Read Pfaffian state, but with the difference that each quasihole operator should now contain both $\sigma^\uparrow$ and $\sigma^\downarrow$ at the same position, coupled to the charge field $e^{i\phi_c/2}$. The correlator of these quasihole operators, together with electronic operators, gives the quasihole wave functions (\ref{eq:QHWaveFunctions}),
\be
\nn\Psi^\text{qh}(\{z^\uparrow_i\},\{z^\downarrow_j\}, \{w_k\}) = ~~~~~~~~~~~~~~~~~~~~~~~~~~~~\\
~~~~~~~\langle\prod_i^{N^\uparrow}\prod_j^{N^\downarrow}\psi_e^\uparrow(z^\uparrow_i)\psi_e^\downarrow(z^\downarrow_j)\prod_{k = 1}^{2n}\psi_{qh}(w_k)\rangle.
\label{eq:qh-wf-ising-cft}
\ee
Note that, similar to quasiholes of single-layer Pfaffian states, in a system without a boundary, the quasiholes appear as pairs. For the coupled Moore-Read state this means that the simplest excitation consists of two pairs of $\sigma^\uparrow\sigma^\downarrow$ operators. As an example, the correlator of two quasihole operators at positions $w_1$ and $w_2$ leads to the following wave function,
\be \nn
\Psi^\text{2qh}(\{z^\uparrow_i\},\{z^\downarrow_j\}, w_1, w_2) =~~~~~~~~~~~~~~~~~~~~~~~~~~~~~~~~~ \\\nn 
\times~ {\rm Pf}\left(\frac{(z^\uparrow_i-w_1)(z^\uparrow_j-w_2) + i\leftrightarrow j}{z^\uparrow_i-z^\uparrow_j}\right)~~~~~~~~~~~~~~~~~\\\nn 
\times ~{\rm Pf}\left(\frac{ (z^\downarrow_i-w_1)(z^\downarrow_j-w_2) + i\leftrightarrow j}{z^\downarrow_i-z^\downarrow_j}\right)~~~~~~~~~~~~~~~~~\\
\times \prod_{i<j=1}^{N_\uparrow}(z^\uparrow_i - z^\uparrow_j)\prod_{i<j=1}^{N_\downarrow}(z^\downarrow_i - z^\downarrow_j)\prod_{i=1}^{N_\uparrow}\prod_{j=1}^{N_\downarrow}(z^\uparrow_i - z^\downarrow_j).~~~
\ee

Since a quasihole operator of the form Eq.~(\ref{qhop}) is the smallest-charge operator that is local with respect to electrons, attempting to break this object further apart by separating  two quasiholes $\sigma^\uparrow$ and $\sigma^\downarrow$ costs energy and results in branch cuts in the electron coordinates of the expression for the quasihole wave function. Correspondingly, sectors that contain unpaired $\sigma^{\uparrow/\downarrow}$ operators are said to be \emph{confined}.  Confined quasiholes can be thought of as being connected to each other (or to the edge of a system with boundary) with physical `strings'. As a consequence, pulling them apart will cost energy so these sectors are naturally gapped out. 

Thus, the lowest-energy quasihole excitations must appear as bound objects, shared by the two layers and occurring at the same position. This can be traced back to the fact that ${\mathcal H}_\text{220}^{\uparrow\downarrow}$  is not a real three-body Hamiltonian and the pairing properties of its ground state are induced by the tunneling term, which in turn binds the quasiholes in the two layers and requires them to occur at the same position. A real-space demonstration of this quasihole binding is presented in Appendix~\ref{apx:exact}.

\subsection{From Ising$^\uparrow \times$ Ising$^\downarrow$ to U(1)$_4$}
We will now further investigate the effects of  the confinement of $\sigma^{\uparrow/\downarrow}$ operators on the topological order of the system. We will see that the topological order is in fact Abelian and we will find an alternate description of the electron and quasihole operators in terms of the primary fields of a U(1) $\times$ U(1) CFT.

\begin{table*}[ht]
\begin{center}
\begin{tabular}{|c|c|c|}
\hline
Confined[Ising$^\uparrow$ $\times$ Ising$^\downarrow$ $\times$ U(1)$_4$] & U(1)$_2$ $\times$ U(1)$_2$ & $h$\\
\hline\hline
$(\psi^\uparrow,1,e^{i\phi_c}) = (1,\psi^\downarrow,  e^{i\phi_c}) = (\psi^\uparrow,\psi^\downarrow,1)$  = (1,1,1)  & (1,1) & $0$  \\ 
\hline
$(1,1,e^{i\phi_c}) = (\psi^\uparrow,1,1) = (1,\psi^\downarrow,1)$ &  $(e^{i\phi/\sqrt{2}},e^{i\phi'/\sqrt{2}})$ & $1/2$ \\
\hline
$\big((\sigma^\uparrow, \sigma^\downarrow)_+, e^{i\phi_c/2}\big) = \big((\sigma^\uparrow, \sigma^\downarrow)_-, e^{-i\phi_c/2}\big)$  & $(e^{i\phi/\sqrt{2}}, 1)$ & $1/4$ \\
\hline
$\big((\sigma^\uparrow, \sigma^\downarrow)_-, e^{i\phi_c/2}\big) = \big((\sigma^\uparrow, \sigma^\downarrow)_+, e^{-i\phi_c/2}\big)$ & $(1,e^{i\phi'/\sqrt{2}})$ &  $1/4$ \\
\hline
\end{tabular}
\caption{The final topological theory describing the coupled Moore-Read state. The fields $\phi$ and $\phi'$ correspond to first and second U(1)$_2$, respectively.}
\label{finalbulk}
\end{center}
\end{table*}

Focusing first on the neutral sector, we start from the Ising$^\uparrow \times$ Ising$^\downarrow$ CFT with the following nine independent sectors,
\be
\label{ising-pure}
\begin{tabular}{|c|c|c|}
\hline
\multicolumn{2}{|c|}{Ising$^\uparrow$ $\times$ Ising$^\downarrow$} & $h$\\
\hline\hline
(1,1) &  & $0$ \\
\hline
($\sigma^\uparrow$,1) & (1,$\sigma^\downarrow$) & $1/16$  \\ 
\hline 
($\sigma^\uparrow$,$\sigma^\downarrow$) & & $1/8$  \\ 
\hline
($\psi^\uparrow$, 1) & (1, $\psi^\downarrow$)   & $1/2$ \\
\hline
($\sigma^\uparrow$, $\psi^\downarrow$) & ($\psi^\uparrow$,$\sigma^\downarrow$) & $3/16$  \\ 
\hline
($\psi^\uparrow$, $\psi^\downarrow$) &  & $1$ \\
\hline
\end{tabular}
\ee
where $h$ denotes the conformal dimension of each sector. The confinement of $\sigma^\uparrow$ and $\sigma^\downarrow$ sectors of Ising$^\uparrow$ $\times$ Ising$^\downarrow$, implies that all sectors that contain unpaired $\sigma^{\uparrow/\downarrow}$ operators, i.e.~$(\sigma^\uparrow, 1), (1, \sigma^\downarrow), (\sigma^\uparrow, \psi^\downarrow)$ and $(\psi^\uparrow, \sigma^\downarrow)$ are also confined. 

Once removing the confined sectors, we see that the remaining five sectors in (\ref{ising-pure}) can no longer be considered topologically distinct. For example, in Ising$^\uparrow \times$ Ising$^\downarrow$ the bosonic $(\psi^\uparrow, \psi^\downarrow)$ can be distinguished from the identity (1,1) because they have non-trivial monodromy with $(1, \sigma^\downarrow)$, $(\sigma^\uparrow, 1)$, etc. In the absence of these sectors, (1,1) and $(\psi^\uparrow, \psi^\downarrow)$ are indistinguishable by braiding and we identify these topological sectors. Similarly, $(1, \psi^\downarrow)$ and $(\psi^\uparrow, 1)$ merge into a single topological sector. Finally, for the fusion rules to be consistent,\cite{bais09a} we see that $(\sigma^\uparrow, \sigma^\downarrow)$ must split into two sectors, which we label $(\sigma^\uparrow, \sigma^\downarrow)_+$ and  $(\sigma^\uparrow, \sigma^\downarrow)_-$. The remaining four independent sectors form a new CFT that is isomorphic to the Abelian U(1)$_4$, as is summarized below.
\be
\begin{tabular}{|c|c|c|}
\hline
Confined[Ising$^\uparrow$ $\times$ Ising$^\downarrow$] & U(1)$_4$ & $h$\\
\hline\hline
($\psi^\uparrow$, $\psi^\downarrow$)  = (1,1)  & 1 & $0$  \\ 
\hline
($\psi^\uparrow$, 1) = (1, $\psi^\downarrow$) &  $e^{i\phi_n}$ & $1/2$ \\
\hline
$(\sigma^\uparrow, \sigma^\downarrow)_+$  & $e^{i\phi_n/2}$ & $1/8$ \\
\hline
$(\sigma^\uparrow, \sigma^\downarrow)_-$ & $e^{-i\phi_n/2}$ &  $1/8$ \\
\hline
\end{tabular}
\label{neutral}
\ee

This type of reduction from one topological order to another has been discussed in some detail in the context of topological phase transitions in Ref.~\onlinecite{bais09a}. There the transition was considered as driven by the condensation of a bosonic sector, with identification and splitting of some sectors and confinement of others as a consequence.  In the case of a system with Ising $\times$ Ising topological order one can imagine the condensation of $(\psi, \psi)$, which would then lead to U(1)$_4$, precisely as is summarized in (\ref{neutral}). In our system there is no evidence for the condensation of $(\psi, \psi)$, nevertheless the same result is obtained from the confinement of single $\sigma$ sectors. A similar analysis in the context of one-dimensional spin chains is studied in Ref.~\onlinecite{mansson13}.

Up to this point our CFT has reduced to two U(1)$_4$ sectors -- one that represents the neutral sector and another for the charge sector. In terms of the primary fields of U(1)$_4$  $\times$ U(1)$_4$ the bosonic electron operators can be defined as,
\be
\label{eop-a}
\nn
\psi_e^\uparrow(z^\uparrow) = \sqrt{2}\cos{\phi_n(z^\uparrow)} e^{i\phi_c(z^\uparrow)},\\
\psi_e^\downarrow(z^\downarrow) = i\sqrt{2}\sin{\phi_n(z^\downarrow)} e^{i\phi_c(z^\downarrow)},
\label{eop-u} 
\ee
and the correlator of these operators also reproduces the coupled Moore-Read state, Eq.~(\ref{eq:doublePfaffian}).\cite{NoteElectronOperators} Note that unlike the charge boson field,\cite{footnote:bg charge} the neutral vertex operators $e^{i\phi_n}$ are not balanced by a background charge. Hence, only terms that balance the overall neutral boson ``charge" contribute to correlators of these operators. Note also that the neutral parts of the electron operators, defined as symmetric and anti-symmetric combinations of $e^{\pm i\phi_n}$ in Eq.~(\ref{eop-u}), correspond to the $(\psi^\uparrow, 1)$ and $(1,\psi^\downarrow)$ sectors of Ising$^\uparrow$ $\times$ Ising$^\downarrow$, which can be viewed as a $\mathbb{Z}_2$ orbifold of U(1)$_4$,\cite{dijkgraaf89} hence recovering the definitions in Eq.~(\ref{eops}).

Similarly, the quasihole operators can be expressed as,
\be
\label{qhop-a}
\psi^+_{qh}(w_1) &=& e^{i\phi_n(w_1)/2}e^{i\phi_c(w_1)/2},\\\nn
\psi^-_{qh}(w_2) &=& e^{-i\phi_n(w_2)/2}e^{i\phi_c(w_2)/2}, 
\ee
which, in the Confined[Ising$^\uparrow$ $\times$ Ising$^\downarrow$ $\times$ U(1)$_4$]  language, are equivalent to 
\be
\psi^+_{qh}(w_1) &=& (\sigma^\uparrow(w_1)\sigma^\downarrow(w_1))_+ e^{i\phi_c(w_1)/2},\\\nn
\psi^-_{qh}(w_2) &=& (\sigma^\uparrow(w_2)\sigma^\downarrow(w_2))_- e^{i\phi_c(w_2)/2}.
\ee
Note that in a system without an edge, any non-vanishing correlator must include both $\psi^+_{qh}$ and $\psi^-_{qh}$, hence these quasiholes always appear as pairs (to satisfy ``charge" neutrality in the neutral sector). 

It is important to note that, despite the fact that we used correlators of the fields of U(1)$_4$ $\times$ U(1)$_4$ to obtain the ground state and quasihole wave functions in Eqs.~(\ref{eop-a}, \ref{qhop-a}), not all 16 sectors of U(1)$_4$ $\times$ U(1)$_4$ lead to valid wave functions. Instead, only those sectors that are local with respect to electron operators, Eq.~(\ref{eop-a}), should be considered to determine the final CFT. Discarding the sectors that are not non-local with respect to electron operators is equivalent to condensing bosonic operators of the form $e^{i\phi_n}e^{i\phi_c}$ within U(1)$_4$ $\times$ U(1)$_4$. The resulting reduced CFT is isomorphic to U(1)$_2$ $\times$ U(1)$_2$, which is the CFT description of the 220 state with 4 distinct sectors, as expected. The correspondence between Confined[Ising $\times$ Ising $\times$ U(1)$_4$] and U(1)$_2$ $\times$ U(1)$_2$ is summarized in Table~\ref{finalbulk}.

\subsection{Analysis of the Edge}
The lowest energy subspace of the effective two-body Hamiltonian $\hat {\mathcal H}^{\uparrow\downarrow}_{220}$ in the presence of extra flux was studied in section~\ref{sec:excitations}. There it was shown that the character of the edge spectrum, for an even total number of particles, is given by Eq.~(\ref{eq:JCCharacter}), i.e.
\be
\nn
\chi^\text{JC} =  [(\chi^\text{MW}_+ )^2 + (\chi^\text{MW}_- )^2] \times \chi^\text{Bose}.
\ee
Here $\chi^\text{Bose}$, which is  the character of a chiral boson [Eq.~(\ref{eq:chiralbosonchi})], represents the charge sector while the neutral sector consists of two parts corresponding to the even/even and odd/odd distributions of particles between the two layers. Interestingly, the expression for the neutral sector is equivalent to the character of the vacuum sector of the compactified chiral boson, U(1)$_4$, i.e.
\be
\label{eq:U14Character}
(\chi^\text{MW}_+ )^2 + (\chi^\text{MW}_- )^2 = \chi^{U(1)_4}_0.
\ee
This can be shown explicitly by noting that $\chi^{U(1)_4}_0$ can be written as,
\be
\label{eq:cagen}
\chi^{U(1)_4}_0 = \chi_0 \times \chi^\text{Bose},
\ee
where
\be
\label{eq:U14IdentitySector}
\chi_0 = \sum_{n=-\infty}^\infty q^\frac{(2 n)^2}{2}
\ee
represents the addition of bosonic descendant fields of the form $e^{\pm i2n\phi}$ to the chiral algebra of the chiral boson in the neutral sector. The full edge spectrum, including both the charge and the neutral sectors, can then be written as,
\be
\chi^\text{JC} = \chi_0 \times \chi^\text{Bose} \times \chi^\text{Bose},
\ee
which corresponds to the character of U(1)$_4$ $\times$ U(1). 

Before proceeding with an explanation of this observation, note that the difference between the characters of the uncompactified U(1) and the compactified U(1)$_4$ is that the chiral algebra of the latter includes bosonic descendant fields of the form $e^{\pm i2n\phi}$ (or their linear combinations) with conformal dimensions $h = 2n^2$ where $n$ is an integer. These operators, which represent creation or annihilation of bosons in the system, create extra modes in addition to the modes of the uncompactified U(1). Therefore, in general, in a system where the number of the underlying particles is fixed, such operators do not enter the chiral algebra and the compactification radius of U(1) does not appear.\cite{footnote1} 

In our two-body Hamiltonian $\hat {\mathcal H}^{\uparrow\downarrow}_{220}$, pairs of particles can freely tunnel between the two layers. Thus, while the total number of particles (or equivalently the total charge) between the two layers is conserved, there is no such restriction on the neutral sector. As a result, the bosonic descendant operators of U(1)$_4$ (or, equivalently the Confined[Ising$^\downarrow$ $\times$ Ising$^\uparrow$]), which forms the neutral part of the CFT, can be added to the chiral algebra of U(1), giving rise to the appearance of one U(1)$_4$ in the edge spectrum. The remaining uncompactified U(1) factor represents the charge sector with fixed total charge, hence the edge spectrum becomes U(1)$_4$ $\times$ U(1).  Eq.~(\ref{eq:U14Character}) is, in fact, a manifestation of the equivalence between the vacuum sector of the $\mathbb{Z}_2$ orbifold of U(1)$_4$  and the combination of the ($\psi^\uparrow$, $\psi^\downarrow$)  and (1,1) sectors of Ising$^\uparrow$ $\times$ Ising$^\downarrow$, as is shown in the first row of (\ref{neutral}). 

Similarly, the lowest energy subspace of ${\mathcal H}_{220}^{\uparrow\downarrow}$ for the case of $N = N_\uparrow + N_\downarrow$ odd, consists of even/odd and odd/even distributions of particles between the two layers. The character of the neutral sector then corresponds to  $\chi^\text{neutral} = 2 \chi^\text{MW}_+\chi^\text{MW}_-$, which is equivalent to the character of the $e^{i\phi_n}$ sector of U(1)$_4$. This represents the equivalence between the $e^{i\phi_n}$ sector of the $\mathbb{Z}_2$ orbifold of U(1)$_4$ and the combination of the (1, $\psi^\downarrow$)  and ($\psi^\uparrow$, 1)  sectors of Ising$^\uparrow$ $\times$ Ising$^\downarrow$, as is shown in the second row of (\ref{neutral}).

As was noted before, a similar study of the edge spectrum of the three-body Hamiltonian ${\mathcal H}_{3\text{-}2}$, Eq.~(\ref{eq:projectiveH}), results in a completely different structure, i.e."SU(2)$_2$ $\times$ SU(2)$_2$. This implies that each layer behaves as an independent Moore-Read state with its own independent charge. However, the ground state of this Hamiltonian at filling fraction $\nu = 1$ is the coupled Moore-Read state, whose electron operators must share the same charge. We have not been able to find a CFT that describes both the ground state and the excitation spectrum of this Hamiltonian, and conjecture that such CFT does not exist. We believe this inconsistency is due to the fact that this Hamiltonian is gapless in the thermodynamic limit.

\section{Conclusions and Outlook}
\label{sec:conclusions}

We have analyzed two model Hamiltonians for coupled quantum Hall bilayers that give rise to ground-state wave functions built from the coupled Moore-Read states, Eq.~(\ref{eq:doublePfaffian}), thus providing a model for the observation of Josephson physics in fractional quantum Hall states.

The first model Hamiltonian, $\hat {\mathcal H}_{3\text{-}2}$, comprises a three-body intra-layer contact interaction and a two-body inter-layer contact interaction. It has the coupled Moore-Read state, Eq.~(\ref{eq:doublePfaffian}), as its ground state and results in an excitation spectrum that is essentially equivalent to that of two independent Moore-Read states. 
More specifically, the zero energy eigenstates of this Hamiltonian can be obtained from those of a system with independent Moore-Read layers by simply multiplying by an inter-layer Jastrow factor, which reflects the inter-layer interaction.  This Hamiltonian has a Goldstone mode reflecting the symmetry that one can move particles freely between layers two at a time and remain in the zero energy subspace.   So long there is no tunneling between the layers, this symmetry remains valid and the corresponding Goldstone mode is present.

The second model Hamiltonian $\mathcal{H}_{220}^{\uparrow\downarrow}$ consists of purely two-body terms, including both interactions and inter-layer tunneling. 
This Hamiltonian is gapped and its ground state is an equal-weight superposition of coupled Moore-Read states for all particle number distributions between the two layers with fixed parity. This state is equivalent to the (gapped) Halperin 220 state in a rotated basis, with the corresponding topological order being U(1)$_2 \times$ U(1)$_2$. 
Correspondingly, while the ground state of this system is a superposition of the ground states of  $\mathcal{H}_{3\text{-}2}$, the excitation spectra of the two systems are quite different, even if we project onto fixed particle numbers in both layers.

Since the $\mathcal{H}_{3\text{-}2}$ Hamiltonian is gapless, it does not have all of the properties one would desire of a true topological phase. For example it is not a correlator of an obvious CFT. Nonetheless, many of its properties are essentially those of the constituent (non-Abelian) Moore-Read states with an excitation spectrum of zero-energy quasihole states given by the product of two Moore-Read spectra, and an edge spectrum given by SU(2)$_2\times$SU(2)$_2$.    However, even infinitesimal tunneling between layers gaps the Goldstone mode, as well as gapping other zero energy states in the presence of quasiholes. The result is a new (and now a proper) topological phase of matter, represented by the Hamiltonian $\mathcal{H}_{3\text{-}2}^\text{JC}(t)$ with nonzero tunneling parameter $t$. The system with tunneling has the same (Abelian) topological order  as $\mathcal{H}_{220}^{\uparrow\downarrow}$. In fact, we found that the ground state of $\mathcal{H}_{3\text{-}2}^\text{JC}(t)$ becomes identical to the ground state of $\mathcal{H}_{220}^{\uparrow\downarrow}$ in the limit of small but nonzero tunneling. 

We demonstrated in detail the connection between the two full quasihole excitation spectra of $\mathcal{H}_{3\text{-}2}^\text{JC}$ and $\mathcal{H}_{220}^{\uparrow\downarrow}$, showing that inter-layer pair tunneling locks the elementary quasiholes of the layers in localized pairs (one quasihole in each layer). In the CFT language, we see that this confinement of quasiholes rules out several sectors of the full SU(2)$_2$ $\times$ SU(2)$_2$ theory (the ones with unpaired elementary quasiholes), which implies that yet more sectors become topologically indistinguishable or invisible - that is, they cannot be distinguished by their braiding properties from any of the remaining, non-confined particles. The introduction of tunneling thus gives a reduction in topological order which is the same as that which would result from topological Bose condensation of the topologically invisible particles -- although no actual condensation in the usual sense happens here. The resulting topological order is identical to that of the Halperin 220 state, given by a U(1)$_2$ $\times$ U(1)$_2$ CFT in the bulk.  

Naively, one might therefore expect that the system exhibits a U(1) $\times$ U(1) edge spectrum when considered at a fixed total number of particles $N$. 
Instead, an interesting twist occurs in the edge spectrum of the coupled Moore-Read state, which takes the surprising form of U(1)$_4 \times$ U(1) at fixed $N$. We attribute this behavior to the fact that in the rotated (symmetric/antisymmetric) basis, the charged and neutral sectors of the 220 state are separated. With the charge quantum number being conserved but without any such conservation restriction on the neutral ``charge", we show that the edge spectrum must take the observed form. 
 
While much of our analysis here has focused on the properties of quasiholes, we note that it is in principle straightforward to construct trial wave functions for quasiparticle states and even for states with quasiholes and quasiparticles. On the one hand, the $220$ state has natural composite fermion excitations. On the other hand, quasiparticle states for  $\hat {\mathcal H}_{3\text{-}2}$ can be constructed from the successful trial wave functions for a single layer Pfaffian considered in Refs.~\onlinecite{Sreejith11,Rodriguez12}. One would expect to find a similar connection between the quasiparticle spectra as for the quasihole spectra, although numerical evidence will be harder to gather, because the quasiparticle states will not be zero energy states of any of the model Hamiltonians.

The physics of a gapless Goldstone mode being gapped by inter-layer tunneling is quite reminiscent of the well-studied physics of the 111 state.\cite{perspectives, Eisenstein111Science, Eisenstein111Nature, MoonYang111, YangMoon111, GunnarSteveBilayer}   Analogous to that case, in the absence of tunneling there is an exact degeneracy associated with moving particles between layers (in the 111 case, particles can be moved between layers one at a time, whereas here they must be moved in pairs).   This degeneracy can also be understood in a different language  where this state can be parametrized in terms of the components of a (pseudo)-spinor, where the direction of the spinor represents the amplitude of the particles in the two layers. Allowing the spinor direction to vary as a function of position, one obtains low energy Goldstone mode excitations (spin waves) of the pseudospin ferromagnet.\cite{SpielmanGoldstone01, GunnarSteveGoldstone}   Analogous to the case of the 111 state, introduction of a tunneling term between the layer breaks the symmetry, fixes the direction of the pseudospin and gaps the Goldstone mode.  In both this case and in the 111 state, the low energy excitations of the system are spin configurations known as merons, which correspond to introducing a quasiparticle vortex in only one of the two spin species. Therefore, independent Majorana-like excitations can form in each layer independently although they are bound together, or confined, at longer distances. The connection to the physics of the 111 state will be explored in more detail in a forthcoming paper.

In addition to the connection to 111 physics, and its associated exciton physics, there are several other connections that would be interesting to explore. One possibility is to consider coupled $\mathbb{Z}_n$ Read-Rezayi wave functions, with tunneling of $n$ particles between layers at a time.   Much of the same confinement physics will remain, although some non-Abelian particles may remain deconfined. Another example to explore would be the tunnel coupling of more than two layers together.  These are also issues that we will defer until a later time.  Much of the discussion here seems somewhat reminiscent of (although not precisely the same as) the work of Ref.~\onlinecite{barkeshli11-orbifold} on orbifold constructions in quantum Hall multilayers.   It would be interesting to explore this connection further.

\begin{acknowledgments}
Discussions with E.~Ardonne, M.~Barkeshli, B.~Estienne and N.~Read are gratefully acknowledged.  The authors would like to thank Nordita and the Aspen Center for Physics their hospitality, and acknowledge support from the Leverhulme Trust under grant ECF-2011-565, the Newton Trust of the University of Cambridge and by the Royal Society under grant UF120157~(G.M.), the European Union under Marie Curie award 299890 QETPM~(L.H.), Science Foundation Ireland principal investigator awards 08/IN.1/I1961 and 12/IA/1697~(J.K.S.) and EPSRC grant EP/I032487/1~(S.H.S.).
\end{acknowledgments}

\appendix

\section{Conventions for states / operators on the sphere}
\label{apx:sphere}

The numerical work presented in this paper was performed for finite systems with spherical geometry. Single-particle orbitals on the plane relate to their counterparts in the spherical geometry via the mapping $z^m \exp(-|z|^2/4)\leftrightarrow u^{S-m}v^{S+m}$ (precisely, states within a spherical droplet on the plane centered around the origin and spanned by the first $N_\phi+1$ orbitals are mapped to the sphere pierced by $N_\phi=2S$ flux quanta). 
We denote wave functions as polynomials in coordinates $z_i$ throughout in the understanding that a corresponding state on the sphere follows via this mapping. To obtain the full form of the many-body wave function on the plane, a factor of $\exp\{-\sum_i |z_i|^2/4\}$ needs to be added.

Generally, quantum Hall Hamiltonians are parametrized by relative angular momenta of particles.\cite{Haldane83} On the sphere, it is more favorable instead to express Hamiltonians in terms of projectors onto (pairs and triplets) of fixed total angular momentum: as the total angular momentum is bounded to $S=N_\phi/2$ on the sphere, states of maximal total angular momentum translate to minimal relative angular momentum.\cite{read96} In a spherical geometry with $N_\phi$ flux quanta, two-body contact interactions are therefore given by projection to the maximal total angular momentum of pairs $M^\text{2-body}_\text{max}=N_\phi$, and we take the following representation to express delta functions in the lowest Landau-level on the sphere
\be
\lambda_2 \sum_{i<j} \mathcal{P}_\text{\tiny LLL}\delta^{(2)}(\mathbf{r}^\uparrow_i - \mathbf{r}^\uparrow_j) \mathcal{P}_\text{\tiny LLL}\simeq \sum_{i<j}P_{ij}(M^\text{2-body}_\text{max}).~~
\ee
An equivalent construction for the three-body terms in (\ref{eq:projectiveH}) includes projectors onto the largest total angular momentum eigenstates for triplets of particles with $M^\text{3-body}_\text{max}=3N_\phi/2$, yielding
\begin{align}
\lambda_3 \sum_{i<j<k}\mathcal{P}_\text{\tiny LLL}\delta^{(2)}(\mathbf{r}^\uparrow_i - \mathbf{r}^\uparrow_j)\delta^{(2)} (\mathbf{r}^\uparrow_j - \mathbf{r}^\uparrow_k) \mathcal{P}_\text{\tiny LLL} \nonumber\\
\simeq \sum_{i<j<k}P_{ijk}(M^\text{3-body}_\text{max}).~~~
\end{align}
We have defined the overall normalization $\lambda_n$ of our Hamiltonians such that the prefactor of projectors onto individual pair/triplet are equal to unity. Projections onto the lowest Landau level are omitted in the main text for brevity.

The explicit form of these projectors can be expressed in terms of creation operators for pairs of particles $\hat \Pi_m^{\dagger}(\uparrow\uparrow,\mathbf{r})$ at a given relative angular momentum $m$. In Eq.~(\ref{eq:VUmklapp}) in the main text, we considered pair creation operators on the plane. Let us make the construction more explicit for the sphere. With the remarks of the preceding paragraph, relative angular momentum $m$ implies total angular momentum $J=2S-m$. A complete basis for the corresponding angular momentum multiplet $|J,M\rangle$ is obtained by coupling the two states $| l_i, m_i\rangle$ which describe the two members of the pair. The matrix elements for the transformation between these bases are given by the Clebsch-Gordon coefficients $C_{l_1,m_1; l_2, m_2}^{J,M} = \langle J,M | l_1,m_1; l_2, m_2 \rangle$, such that 
\begin{align}
\label{eq:}
|J,M\rangle
&= \!\! \sum_{m_1,m_2} \!\! \langle  l_1,m_1; l_2, m_2 | J,M \rangle | S, m_1 \rangle_1 \otimes | S, m_2 \rangle_2\nonumber\\
&= \!\! \sum_{m_1,m_2} \!\! C_{l_1,m_1; l_2, m_2}^{J,M} \hat a^\dagger_{m_1} \hat a^\dagger_{m_2} |\text{vac}\rangle, \nonumber\\
&\equiv \hat \Pi^{\dagger}_{J,M} |\text{vac}\rangle
\end{align}
with creation operators $ \hat a^\dagger_{m}$ for a particle with $L_z=m$, and equivalently defining creation operators for pairs $\hat \Pi^{\dagger}_{J,M}$. (Non-zero contributions arise only for $m_1+m_2=M$). Hence, the projector has the second quantized form
\be
\label{eq:projectors_sphere}
\sum_{i<j} P_{ij}(J) = \sum_M  \hat\Pi^{\dagger}_{J,M} \hat\Pi_{J,M}
\ee
These expressions generalize straightforwardly to the case with spin, defining
\be
\label{eq:pair_creation}
\hat \Pi^{\dagger}_{J,M}(\sigma_1, \sigma_2)  = 
\sum_{m_1,m_2} \!\! C_{l_1,m_1; l_2, m_2}^{J,M} \hat a^\dagger_{m_1,\sigma_1} \hat a^\dagger_{m_2,\sigma_2}.
\ee

\section{Mapping between $\Psi_0(t)$ and $\Psi_{220}$}
\label{apx:mapping220}

\subsection{Wave functions}

In this appendix, we show that the Halperin $220$-state is precisely a superposition of coupled Moore-Read wave functions with different numbers of spin-up and spin-down bosons, $\Psi_0^{N_\uparrow, N_\downarrow}$. The underlying idea is to perform a basis transformation from spin-up and spin-down eigenstates $|\sigma\rangle$ to their symmetric and antisymmetric superpositions (\ref{eq:BasisChange}), $|\pm\rangle = 1/\sqrt{2} (\state{\uparrow}\pm\state{\uparrow}) $.
As we do not know the precise superposition of $\Psi_0^{N_\uparrow, N_\downarrow}$ that yields the 220 state in this basis, let us proceed in reverse and start by writing the $220$-state in the $|\pm\rangle$ basis:
\begin{align}
\Psi_{220} = &\prod_{i<j}^{N/2} (z^+_i-z^+_j)^2 \prod_{i<j}^{N/2} (z^-_i-z^-_j)^2 \nonumber\\
 =& \Psi_{111}\Pf \frac{1}{z_i^+ - z_j^- },
\end{align}
which we have written as a paired state\cite{GunnarSteveBilayer} using Cauchy's identity $\Psi_{001}\Pf [1/(z_i^+ - z_j^- )] = \Psi_{110}$. Note that the $111$-state can be written as $\Psi_{111} = \prod (z_i-z_j)$, meaning that it has identical correlations between any two particles, so it is not necessary to indicate spin degrees of freedom explicitly. In particular, the state has the same form in any (pseudo-)spin basis. We follow Ho\cite{Ho95} to denote the spin and spatial coordinates separately in the pair wave-function, and adopt the notation
\begin{align}
\Psi_{220} =& \Pf \left[ \frac{|+ - \rangle}{z_i - z_j} \right] \prod_{i<j}^{N}(z_i-z_j) \nonumber\\
=& 2^{-\frac{N}{2}}\Pf \left[  \frac{|+ - \rangle + |- + \rangle}{z_i - z_j} \right]  \prod_{i<j}^{N}(z_i-z_j) 
\\ \equiv & 2^{-\frac{N}{2}} \Pf \left[  \frac{\mathbf{e}_z  \cdot (i \boldsymbol{\hat\sigma} \sigma_y)_{\alpha\beta}}{z_i - z_j} \right]  \prod_{i<j}^{N}(z_i^\alpha-z_j^\beta), \nonumber
\label{eq:220_dvector}
\end{align}
where kets $|\alpha \beta\rangle$ indicate the spin-states of the two members of a Cooper pair, and their coordinates are written as $z_i$, or $z_i^\alpha$ when a specific spin state is represented. 
In the second step, we have chosen to symmetrize the notation, and finally, we use the notation of the symmetric spin-triplet wave function $\chi_{\alpha\beta}$ in terms of the $\mathbf{d}$-vector, $\chi_{\alpha\beta} = \mathbf{d} \cdot (i \boldsymbol{\hat\sigma} \sigma_y)_{\alpha\beta}$, with $\mathbf{d}=\mathbf{e}_z$.
 Now, we take the inverse basis transformation to the original basis of pseudospin up/down. The spin-state takes the form
\begin{align}
|+ - \rangle + | - + \rangle &= \frac{1}{2}
\bigl[ (\spinu_1+\spind_1)(\spinu_2-\spind_2) \nonumber\\
& \quad + (\spinu_1-\spind_1)(\spinu_2+\spind_2) \bigr]\nonumber\\
&=|\uparrow \uparrow\rangle - |\downarrow \downarrow\rangle
\end{align}
In the language of spin-triplet pairing, this corresponds to $\mathbf{d}=-\mathbf{e}_x$, i.e., the transformation amounts to a $\pi/2$ rotation of the spin reference frame around the $y$-axis.
Replacing the pair correlation function in (\ref{eq:220_dvector}) accordingly, and using the explicit definition of the Pfaffian of an $n\times n$ matrix $M_{i,j}$,
\be
\label{eq:def_Pfaffian}
 \Pf M = \frac{1}{2^\frac{n}{2}(\frac{n}{2})!}\sum_{\sigma \in S_n} \text{sgn}(\sigma)  \prod_{k = 1}^{n/2} M_{ \sigma(2k-1), \sigma(2k)},~~
\ee
where $\sigma$ are elements of the permutation group $S_n$, we find the explicit expression
\begin{align}
\nn\Psi_{220} = \Psi_{111} \frac{1}{2^N (\frac{N}{2})! } \sum_{\sigma \in S_N} \text{sgn}(\sigma)  \prod_{k = 1}^{N/2}\left[ \frac{|\uparrow \uparrow\rangle - |\downarrow \downarrow\rangle}{z_{\sigma(2k-1)} - z_{\sigma(2k)}} \right]
\end{align}
The pair wave function states that each pair is either both spin-up or both spin-down (with a minus-sign). As we sum over permutations, we can make those choices explicit for all terms up to reordering of the permutation. The number of choices to be made is equal to the binomial coefficient, and we have
\begin{align}
\nn\Psi_{220} = &\frac{ \Psi_{111}}{2^N (\frac{N}{2})! } \sum_{\sigma \in S_N} \text{sgn}(\sigma)  \sum_{p=0}^{N/2}  \binom{N/2}{p}\nn\\ 
&\times \prod_{k = 1}^{p}\left[ \frac{1}{z^\uparrow_{\sigma(2k-1)} - z^\uparrow_{\sigma(2k)}} \right]\nn\\
&\times \prod_{k' = 1}^{N/2-p}\left[ \frac{-1}{z^\downarrow_{\sigma(2p+2k'-1)} - z^\downarrow_{\sigma(2p+2k')}} \right] \nn\\
= &\Psi_{111}\sum_{p=0}^{N/2} \sum_{\sigma \in S_N} \text{sgn}(\sigma)  \frac{1}{2^{2p}p!}\prod_{k = 1}^{p}\left[ \frac{1}{z^\uparrow_{\sigma(2k-1)} - z^\uparrow_{\sigma(2k)}} \right]\nn\\
&\times \frac{(-1)^{(\bar p)}}{2^{2\bar p}(\bar p)!} \prod_{k' = 1}^{\bar p}\left[ \frac{1}{z^\downarrow_{\sigma(2p+2k'-1)} - z^\downarrow_{\sigma(2p+2k')}} \right],
\end{align}
where we have used the shorthand notation $\bar p = N/2-p$. We can now identify $N_\uparrow = 2p$, and $N_\downarrow = 2\bar p = N-N_\uparrow$, as well as noting that the products of up-spin pair wave functions forms a complete Pfaffian (and similarly for the down-spin part). Hence, we see that the preceding expression is precisely the superposition
\be
\Psi_{220} = \mathcal{S}_{\uparrow\downarrow}\left[(-1)^{N_\downarrow}\Psi_0^{N_\uparrow, N_\downarrow}\right],
\ee
where $\mathcal{S}_{\uparrow\downarrow}$ is the operator that symmetrizes over all possible assignments of up-spin and down-spin to the particles.

\iffalse
\subsection{CFT operators}
The basis change is explained more easily in terms of the CFT representation of fermion fields. Let us focus only on the neutral part of the Fermion operators, only, involving a neutral Boson field denoted $\phi_n\equiv \phi$ within this section. We then have the representations as in Eq.~(\ref{eop-a}) (up to overall charge factors),
\begin{align}
\Psi_{\uparrow(\downarrow)} \simeq e^{i\phi} + (-) e^{-i\phi}.
\end{align}
Hence, the (anti-)symmetric superpositions yield
\begin{align}
\Psi_{\pm} = (\Psi_\uparrow \pm \Psi_\downarrow) \simeq e^{\pm i \phi}.
\end{align}
The (neutral part of the) $220$-state has the form 
\begin{align}
\Psi_{220} \simeq & \langle \Psi_+(z_1) \cdots \Psi_+(z_{N/2}) \Psi_-(z_{N/2+1}) \cdots \Psi_-(z_{N}) \rangle \nn\\
	=&\langle \underbrace{(\Psi_\uparrow + \Psi_\downarrow)|_{z_1}\cdots}_{N/2 \text{ times} } \underbrace{(\Psi_\uparrow - \Psi_\downarrow)|_{z_{N/2+1}}\cdots}_{N/2 \text{ times} }\rangle \nn\\
	=& \langle \underbrace{(\Psi_\uparrow(z_1)\Psi_\uparrow(z_{N/2+1}) - \Psi_\downarrow(z_1)\Psi_\downarrow(z_{N/2+1}) +  \Psi_\uparrow(z_1) \Psi_\downarrow(z_{N/2+1} - \Psi_\downarrow(z_1) \Psi_\uparrow(z_{N/2+1})\cdots}_{N/2 \text{ times} }\rangle
\end{align}
{\bf (** needs to be fixed **)} Again, contributions to the wave function arise by choices of pairs with either spin-up or spin-down, the pair-wave function from the OPE of pairs of $\Psi$-fields yields a Pfaffian for each layer, and sectors with $\Delta N_\uparrow = 2$ differ by an overall sign, as above.
\fi

\subsection{Hamiltonian}
Let us now derive the parent Hamiltonian for the coupled Moore-Read state, starting from the known parent Hamiltonian for the $220$-state given in (\ref{eq:Ham220}), and we express the pseudopotentials in terms of the pair creation / annihilation operators (\ref{eq:pair_creation}), so the Hamiltonian can be written in brief as
\be
\hat {\mathcal H}_{220} = \hat\Pi^\dagger_{++} \hat\Pi_{++} + \hat\Pi^\dagger_{--} \hat\Pi_{--},
\ee
and we imply the relative angular momentum to be zero, as well as summation over the angular momenta $M$ as per Eq.~(\ref{eq:projectors_sphere}). The Hamiltonian in the basis of spins $\uparrow,\downarrow$ is obtained by making the replacements
\be
\hat a_{\pm} = \frac{1}{\sqrt{2}}(\hat a_\uparrow \pm \hat a_\downarrow).
\ee
Shortening the notation for Clebsch-Gordon coefficients to include only the angular momentum indices $C_{l_1,m_1; l_2, m_2}^{J,M} \equiv C_{m_1m_2}$, and introducing the shorthand $(\sigma_1\sigma_2\sigma_3\sigma_4) \equiv \hat a^\dagger_{m_1,\sigma_1} \hat a^\dagger_{m_2,\sigma_2} \hat a_{m_3,\sigma_3}\hat a_{m_4,\sigma_4}$, we have
\begin{align}
\hat {\mathcal H}_{220} = &\frac{1}{2}\sum C_{m_1m_4}C^*_{m_2m_3}\bigl[ (\uparrow\uparrow\uparrow\uparrow) + (\downarrow\downarrow\downarrow\downarrow) \nn\\
&+ (\uparrow\downarrow\uparrow\downarrow) + (\downarrow\uparrow\downarrow\uparrow) + (\uparrow\downarrow\downarrow\uparrow) + (\downarrow\uparrow\uparrow\downarrow) \nn \\
&+ (\uparrow\uparrow\downarrow\downarrow) + (\downarrow\downarrow\uparrow\uparrow) \bigr].
\end{align}
Here, we can identify the terms of the parent Hamiltonian of Eq. (\ref{eq:Ham220UD}), as the intra-layer contact repulsions $\hat V_0^{\uparrow\uparrow} + \hat V_0^{\downarrow\downarrow}$ (first line), inter-layer contact repulsion $\hat V_0^{\uparrow\downarrow}$ (second line) and the local pair tunneling terms $\hat V_0^\text{tun}$ (third line).

\section{Coupled Pfaffian as the exact ground state of $\hat {\mathcal H}^{\uparrow\downarrow}_{220}$}
\label{apx:exact}
In this appendix we show explicitly that the coupled Moore-Read state is the exact zero-energy ground state of $\hat {\mathcal H}^{\uparrow\downarrow}_{220}$, or equivalently, of $\hat {\mathcal H}^\text{eff}(\alpha=1)$. We first focus on the example of four particles and then generalize the result. The effective two-body Hamiltonian with pair tunneling can be written using the following notation,
\be
\nn\hat {\mathcal H}^\text{eff}(\alpha)&=& \sum_{i,j = 1}^N \delta^{(2)}(z_i - z_j)
\sum_{s\neq s' = \uparrow, \downarrow}
\big(|ss\rangle\langle ss| +  |ss'\rangle\langle  ss'| \\
&+&  |ss'\rangle\langle s's| +~\alpha |ss\rangle\langle s's'|\big)_{ij},~~~~~~~~~~~~~~~~~~~~
\ee
where we have inserted the parameter $t$ to represent the strength of the pair tunneling term. Starting with a total of four particles, the coupled Moore-Read wave function consists of three sectors: either all particles are in the top layer, all particles in the bottom layer or they divide equally between the two layers. Using the notation $(ij) \equiv (z_i - z_j)$, we start with an (almost) general superposition of these sectors (assuming equal weight for the sectors $(N_\uparrow,N_\downarrow)=(4,0)$, and $(0,4)$, for ease of writing), which can be written as follows,
\be 
\nn|\Psi\rangle &=& \Psi_0^{4,0} + k \Psi_0^{2,2} + \Psi_0^{0,4} \\ \nn
&=& \Big((13)(14)(23)(24) - (12)(14)(23)(34) \\\nn
&+& (12)(13)(24)(34)\Big)|s~s~s~s\rangle\\\nn
&+& k\Big((13)(14)(23)(24)|s~s~s's'\rangle\\\nn
&-& (12)(14)(23)(34) |s~s's~s'\rangle\\
&+& (12)(13)(24)(34)|s~s's's\rangle\Big).
\ee
Here we have omitted $\sum_{s\neq s' = \uparrow, \downarrow}$ and the parameter $k$ is the relative weight of the sectors with two particles in each layer, with respect to sectors where all four particles are in one layer. Applying the effective Hamiltonian on this wave function, the terms that are not trivially zero take the form,\cite{projection}
\begin{align}
&\nn\hat {\mathcal H}^\text{eff}(\alpha)|\Psi\rangle 
\\\nn= &\Big(\delta^{(2)}(z_1 - z_2) + \delta^{(2)}(z_3 - z_4)\Big) (13)(14)(23)(24)
\\\nn
&\times~\Big((1 + \alpha k)(|s~s~s~s\rangle +(k + \alpha)|s~s~s's'\rangle \Big)~~~~~~~~
\\\nn 
&-~\Big(\delta^{(2)}(z_1 - z_3) + \delta^{(2)}(z_2 - z_4)\Big) (12)(14)(23)(34)
\\\nn
&\times~\Big((1 + \alpha k)(|s~s~s~s\rangle +(k + \alpha)|s~s's~s'\rangle \Big)~~~~~~~~
\\\nn
&+~\Big(\delta^{(2)}(z_1 - z_4) + \delta^{(2)}(z_2 - z_3)\Big)  (12)(13)(24)(34)
\\
&\times~\Big((1 + \alpha k)(|s~s~s~s\rangle +(k + \alpha)|s~s's~s'\rangle \Big).~~~~~~~
\end{align}
Thus, we can solve the model for its ground state for $|\alpha|=1$, by setting $k = -\alpha = \pm 1$ we see that the remaining terms vanish and $|\Psi\rangle$ is found to be the exact, zero-energy ground state of $\hat {\mathcal H}^\text{eff}(\alpha=1)\equiv \hat {\mathcal H}^{\uparrow\downarrow}_{220}$. (Generally, alpha plays the role of a phase difference of different particle number sectors, so it could equivalently be chosen as a complex number of unit norm -- the problem is still solvable in that case, with correspondingly adjusted relative phases in the components of the wave function.) 

In general, if we denote the ground state as,
\be 
\nn|\Psi\rangle &=& \sum_{s\neq s' = \uparrow, \downarrow}\sum_{n = 0:2}^N k_n M_n M_{N-n} \prod_{i<j = 1}^{N} (z_i - z_j)\\ &\times& ~~|\text{P}(N_s = n, N_{s'} = N-n)\rangle
\ee
where $M_n$ is the Pfaffian of $n$ particles,
\be
\nn M_n &=& \frac{1}{2^{n/2}(n/2)!}\sum_{\sigma \in S_n} \text{sgn}(\sigma)~~~~~~~~~~~~~~~\\
&\times& \prod_{k = 1}^{n/2} \frac{1}{z_{\sigma(2k-1)} - z_{\sigma(2k)}},
\ee
$k_n$ is the amplitude of the corresponding term, and $ \text{P}(N_s = n, N_{s'} = N-n)
$ denotes all permutations of $n$ particles in the $s$ layer and $N-n$ particles in the $s'$ layer, then it can be shown that $\hat {\mathcal H}^{\uparrow\downarrow}_{220}|\Psi\rangle = 0$ if 
a recursive equation of the form 
\be
2k_n  = -\alpha(k_{n-2} + k_{n+2}),
\ee
\\
with boundary conditions, $k_{N-n} = k_n$ is satisfied. This equation always has solutions of the form $k_n =\pm 1$, $\alpha = -1$, and $k_n = - k_{n\pm2} = \pm 1$ with $\alpha = 1$, therefore $|\Psi\rangle$ is an exact, zero-energy ground state of $\hat {\mathcal H}^{\uparrow\downarrow}_{220}$, or $\hat {\mathcal H}^\text{eff}(\alpha=\pm1)$.

A similar analysis can be done to show that in the presence of extra flux, quasiholes are created at the same position in both layers. The example of two quasiholes and four particles might be illuminating. Suppose the particles in each layer see the quasiholes at different positions, for example, let the particles in the $s$ layer see the quasiholes at positions $w_1$ and $w_2$ while the particles in the $s'$ layer see them at positions $w'_1$ and $w'_2$. Using the notation $\beta_{ij} = (z_i - w_1)(z_j - w_2) + (z_j\leftrightarrow z_i)$ and $\beta'_{ij} = (z_i - w'_1)(z_j - w'_2) + (z_j\leftrightarrow z_i)$, the expression for the quasihole wave function takes the form,
\be 
\nn|\Psi\rangle^\text{qh} &=& \Big((13)(14)(23)(24) \beta_{12} \beta_{34}
\\\nn
&-& (12)(14)(23)(34)  \beta_{13} \beta_{24}
\\\nn
&+& (12)(13)(24)(34) \beta_{14} \beta_{23}\Big)|s~s~s~s\rangle
\\\nn
&+& k\Big((13)(14)(23)(24)\beta_{12} \beta'_{34}|s~s~s's'\rangle
\\\nn
&-&(12)(14)(23)(34) \beta_{13} \beta'_{24}|s~s's~s'\rangle
\\
&+&(12)(13)(24)(34)\beta_{14} \beta'_{23}|s~s's's\rangle\Big).
\ee
Similar to the previous case, applying the effective Hamiltonian on this wave function results in the following non-trivial terms,
\be
\nn\hat {\mathcal H}^\text{eff}|\Psi\rangle^\text{qh} = \Big(\delta(z_1 - z_2) + \delta(z_3 - z_4)\Big) (13)(14)(23)(24)
\\\nn
\times~\Big(\beta_{12}(\beta_{34} + \beta'_{34}\alpha k)|s~s~s~s\rangle +
\beta_{12}(\beta_{34}k + \beta'_{34}\alpha)|s~s~s's'\rangle \Big)
\\\nn 
-~\Big(\delta(z_1 - z_3) + \delta(z_2 - z_4)\Big) (12)(14)(23)(34)
\\\nn
\times~\Big(\beta_{13}(\beta_{24} + \beta'_{24}\alpha k)|s~s~s~s\rangle +\beta_{13}(\beta_{24}k + \beta'_{24}\alpha)|s~s's~s'\rangle \Big)
\\\nn
+~\Big(\delta(z_1 - z_4) + \delta(z_2 - z_3)\Big) (12)(13)(24)(34)
\\\nn
\times~\Big(\beta_{14}(\beta_{23} + \beta'_{23}\alpha k)|s~s~s~s\rangle +\beta_{14}(\beta_{23}k + \beta'_{23}\alpha)|s~s's~s'\rangle \Big)
\ee
and for $k = -\alpha = \pm1$, we see that for these terms to vanish we must have $\beta_{ij} = \beta'_{ij}$, i.e.~the quasiholes must be at exactly the same position  in both layers.
 
\section{Review of the edge counting of a single Moore-Read State}
\label{apx:mr}
To make our presentation self-contained, we briefly review how to count the angular momentum degeneracy associated with the quasiholes of the single-layer Moore-Read state. In Ref.~\onlinecite{read96}, Read and Rezayi start with the explicit form of the quasihole wave functions for the Moore-Read state, as per their Equation 2.14:
\begin{align}
\label{eq:ReadRezayiQH}
&\Psi^{\text{qh}, \nu=1/q}_{\text{MR},m_1,\ldots m_p}(z_1,\ldots z_N; w_1,\ldots w_{2n}) 
\\\nn = &\frac{1}{2^\frac{N-p}{2}\left(\frac{N-p}{2}\right)!} \prod_{i<j}(z_i-z_j)^q \sum_{\sigma\in S_N} \text{sgn}(\sigma)\\
\times & \prod_{k=1}^p z_{\sigma(k)}^{m_k}\prod_{l=1}^{\frac{N-p}{2}} 
 \frac{ \Phi \bigl(z_{\sigma(p+2l-1)},z_{\sigma(p+2l)} ; w_1, \ldots w_{2n} \bigr)}{z_{\sigma(p+2l-1)} - z_{\sigma(p+2l)} }. \nn
\end{align}
Here $q$ is the number of flux quanta attached to the underlying particles to composite fermionize them. In this paper we focus on bosons with $q=1$. $\Phi$ is a polynomial of the form,
\be
\label{eq:phi}
\Phi \bigl(z_{\sigma(p+2l-1)},z_{\sigma(p+2l)} ; w_1, \ldots w_{2n} \bigr) 
= \frac{1}{(n!)^2}~~~~~~~~~~~~~
\\\nn
\times \sum_{\tau\in S_{2n}}
\prod_{r = 1}^n(z_{\sigma(p+2l-1)} - w_{\tau(2r-1)})(z_{\sigma(p+2l)} - w_{\tau(2r)}),
\ee
which is symmetric under the exchange of $w$'s. 

Read and Rezayi demonstrate that the degeneracy arising with the addition of $n$ flux quanta to the ground state of the system is characterized by two features: firstly, by the number $2n$ and position of quasihole coordinates $\{w_i\}$, $i=1,\ldots 2n$ and, secondly, by the state of $0\le p\le n$ fermions, which can be left unpaired at zero energetic cost, whenever quasiholes are present. 

While the orbital degeneracy would also be found in simple Abelian states, the second contribution represents the characteristic topological degeneracy associated with the non-Abelian nature of the quasihole excitations.  In fact, the physics of the unpaired fermions can be used not only to count excitations, but also to provide signatures for the $p$-wave pairing of composite fermions in the ground state of realistic two-body Hamiltonians in the second Landau-level.\cite{Moller11} 

The state (\ref{eq:ReadRezayiQH}) describes $p$ fermions that are left unpaired. For each flux quantum added to the system, that is one for each \emph{pair} of quasiholes, the unpaired electrons gain an additional degree of freedom, which is analogous to an effective Landau-level orbital that they may occupy. This is one orbital for each flux added above the ground state. We label the states of unpaired fermions with integers $\{m_k\}$, $k = 1,\ldots p$, which represent the orbital that is occupied by a fermion, and we choose $0\le m_1 < m_2 < \ldots <m_p\le n-1$.\cite{read96} 

Hence, for a situation with $p$ unpaired fermions and flux $N_\phi = (N-1)-1+n$, i.e.~$n$ flux quanta above the ground state, there are
\be
\label{eq:DTopo}
d_\text{topo}(n,p) = \binom{n}{p}
\ee
degenerate states with fixed quasihole positions. This counting can be thought of as arising from $p$ fermions living in $n$ orbitals. The overall topological degeneracy adds up to 
\be
D_\text{topo} = \sum_{\substack{p = N\: \text{mod}\: 2 \\ \{N-p\: \text{mod}\: 2 = 0\}}}^n
d_\text{topo}(n,p) = 2^{n-1}~~~~
 \ee
states, where the sum goes only over values of $p$ that match the parity of $N$, so that $N-p$ is always even. This result matches the non-Abelian nature of the quasiholes, which have quantum dimension $d_\text{qh}=\sqrt{2}$.

The orbital degeneracy follows from expanding the quasihole states in terms of symmetric polynomials of the $2n$ quasihole coordinates $\{w_i\}$, as shown in Eq.~\ref{eq:phi}. Read and Rezayi show that the relevant polynomials have a degree of at most $(N-p)/2$, i.e.~the number of unbroken pairs. Hence, this contribution can be thought of as the degeneracy associated with placing $2n$ bosons in a Landau level with $(N-p)/2$ flux quanta, which yields the orbital degeneracy of,
\be
\label{eq:DOrb}
d_\text{orb}(N,n,p) = \binom{(N-p)/2 + 2n}{2n},
\ee
and the total degeneracy of quasihole states of the Moore-Read state for fixed $N$ and $n$ becomes,
\be
D_\text{MR}(N,n) = \sum_{\substack{p = N\: \text{mod}\: 2 \\ \{N-p\: \text{mod}\: 2 = 0\}}}^n
d_\text{orb}(N,n,p)\: d_\text{topo}(n,p).~~~~~~
\ee

Here, we are also interested in the degeneracy of quasihole states as a function of angular momentum. In order to obtain this dependence, we can use Read and Rezayi's analogy of unpaired fermions and bosons filling the respective numbers of orbitals as discussed above. In order to obtain the angular momentum decomposition of each of these terms individually, we can use Euler's generating function 
\be
\mathcal{Z}(N_\text{orb},q,x) = \prod_{m=1}^{N_\text{orb}} \frac{1}{1-x q^m},
\label{eq:partitionFinite}
\ee
to be taken as an infinite series in the abstract variable $q$, and we have introduced an additional parameter $x$.

The degree $m$ of individual powers of $q$ encodes the angular momentum of a corresponding Landau level orbital $z^m$. The parameter $x$ allows us to distinguish between terms stemming from unoccupied orbitals with trivial factors $`1$' and occupied orbitals that are proportional to $x$. For $x=1$, the ensuing series provides the character of a chiral boson field, and powers in $x$ provide additional information on the number of occupied orbitals in the individual terms at fixed angular momentum. This allows us to read off the finite size counting of the number of states with angular momentum $L_z=l$ for $k$ bosons in $N_\text{orb}$ orbitals by taking certain derivatives,
\be
\label{eq:dbose}
d_\text{Bose}(k, N_\text{orb},l) = \frac{1}{l!} \frac{\partial^l}{\partial q^l}\left[ \frac{1}{k!} \left.\frac{\partial^k}{\partial x^k} \mathcal{Z}(N_\text{orb},q,x) \right]\right|_{x,q\to 0}~~
\ee
The analogous counting for fermions follows from mapping the problem of $k$ fermions in $N_\text{orb}$ orbitals onto the corresponding Bose problem where $(k-1)$ orbitals are removed due to Pauli blocking, and hence,
\be
\label{eq:dfermi}
d_\text{Fermi}(k, N_\text{orb},l) = d_\text{Bose}(k, N_\text{orb}-k+1,l).
\ee

Using Read-Rezayi's insights, we obtain the count of quasihole states at a given angular momentum $L_z$ by following the rules of angular momentum addition and convoluting the Bose and Fermi countings. For the Moore-Read Pfaffian state, with $N$ electrons and $n$ additional flux quanta, the $L_z$ dependent degeneracies become,
\be
\label{eq:dMR}
d_\text{MR}(N,n,L_z)  = 
\sum_{\substack{p=N\text{mod} 2 \\ \{N-p\: \text{mod}\: 2 = 0\}}}^{n}
\sum_{l = 0}^{L_z} ~~~~~~~~~~~~~~\\
 d_\text{Bose}\left(2n, \frac{N-p}{2},l\right) d_\text{Fermi}(p,n-1,L_z - l).\nn
\ee
Finally, the degeneracies of state per $L_z$ sector can be translated into the count of multiplets in the total angular momentum $L$, as given in the original paper by Read and Rezayi.  Because eigenstates of rotationally invariant Hamiltonians always occur in $2L+1$-fold degenerate angular momentum multiplets $|L,m\rangle$, with $m\in \{-L,\ldots,L\}$, the number of multiplets $\mu_\text{MR}(L)$ at a given total angular momentum $L$ is given by
\begin{align}
\mu_\text{MR}(N,n,L) = & \:  d_\text{MR}(N,n,L_z = L) \nn\\
&- d_\text{MR}(N,n,L_z = L + 1).
\end{align}
In tables \ref{tab:DegMRxMR} and \ref{tab:DegEffectiveH}, we denote the structure of the zero-energy Hilbert subspaces $\mathcal{L}$  via their multiplet structure using the notation 
\be 
\label{eq:low_energy_hilbert}
\mathcal{L} = \bigoplus_{L=0}^{L_\text{max}}  L^{\mu(L)}.
\ee

Going beyond the counting of quasihole states in finite size systems, the large angular momentum part of the count of quasihole states maps onto the counting of edge states of the system, which is \emph{universal}.\cite{WenEdge90, moore91, WenEdgeReview} 
Formally, the counting of edge modes can be deduced from $d_\text{MR}(N,n,L_z)$ in the limit of a `large correlation hole' with both $n \to \infty $ and $N\to \infty $ (while maintaining $n<N$).\cite{schoutensEdge} In this picture, the largest angular momentum $L_z^\text{max}$ of the quasihole states on a sphere maps to the edge state of a disc with momentum $\Delta m=0$, while general states obey $\Delta m(L_z)=L_z^\text{max}-L_z$. We use this correspondence in this paper to compare finite size data to the edge state counting in the infinite system, which is conveniently described by the characters of conformal field theories.\cite{moore91}

For the Moore-Read state, the edge of the infinite size droplet is described by a product of the Majorana-Weyl character $\chi^\text{MW}$ and the character of a chiral boson $\chi^\text{Bose}$.\cite{MilovanovicRead} The edge spectrum carries information about the parity of the fermion number and hence differs in sectors of even/odd number of particles, with the respective Majorana-Weyl characters,
\be
\label{eq:mwchi}
\chi^\text{MW}_\pm  = \frac{1}{2}\left[ \prod_{m=0} (1 + q^{m + \frac{1}{2}} ) \pm  \prod_{m=0} (1 - q^{m + \frac{1}{2}} ) \right].~~~~~
\ee
Defining the character of the chiral boson as,
\be
\chi^\text{Bose}=\lim_{N_\text{orb}\to\infty} \mathcal{Z}(N_\text{orb},q,1),
\label{eq:chiralbosonchi}
\ee
the character for the Moore-Read edge becomes,
\be
\label{eq:SingleMRCharacter}
\chi^\text{MR}(N) = \chi^\text{B} \times \chi^\text{MW}_{(-1)^N},
\ee
where only the parity of the number of particles in the droplet affects the result.


\begin{references}
\bibitem{NoteTunnelingSign} We assume $t>0$ in this text; $t<0$ implies phase differences preferentially tend to be zero and the resulting state has $\mathbf{d}$-vector $\propto \mathbf{e}_y$.
\bibitem{NoteHEff} By setting $\alpha = 2\pi\epsilon$, the Hamiltonian (\ref{eq:HamEffAlpha}) recovers a pair tunneling term proportional to $2\pi\epsilon$ in analogy with Ref.~\onlinecite{prl12}, where $\epsilon$ is related to the deviation from flux density $n_\phi=1/2$ in the Hofstadter spectrum, i.e., $\epsilon = |n_\phi-1/2|$. We had previously documented a large overlap of the ground state wave function near $\epsilon\simeq 0.16$. Our derivation in section \ref{sec:JosephsonCoupling}  demonstrate that the wave function is indeed exact at $\epsilon\simeq 1 / (2\pi)$, or $\alpha=1$, where the Hamiltonian reduces to (\ref{eq:Ham220UD}).
\bibitem{footnote:oddN} Similarly, the edge theory for odd $N = N^\uparrow + N^\downarrow$ is described by the overall character, $\chi^\text{JC} = 2 \:\chi_+^\text{MW}\chi_-^\text{MW}\chi^\text{Bose}$.
\bibitem{footnote:bg charge} Here and in the following equations we assume that the neutrality condition on charge vertex operators is satisfied by using a background negative charge, see e.g. Ref~\onlinecite{moore91}.
\bibitem{NoteElectronOperators} Note that one could equivalently define the electron operators as $\psi_e^\uparrow(z^\uparrow) = e^{i\phi_n(z^\uparrow)} e^{i\phi_c(z^\uparrow)}$ and $\psi_e^\downarrow(z^\downarrow) =  e^{-i\phi_n(z^\downarrow)} e^{i\phi_c(z^\downarrow)}$, and then further symmetrize the resulting wave function over pairs of particles in each layer. The definitions in~Eq.~(\ref{eop-u}) automatically do this.
\bibitem{footnote1} This is analogous to the fact that in the edge spectrum of Laughlin states, in systems with fixed numbers of particles, the compactification radius of $U(1)$ does not appear and the mode counting corresponds to that of a an uncompactified $U(1)$.
\bibitem{projection} Here, strictly speaking, we are applying the lowest Landau level (LLL) projection of the effective Hamiltonian, ${\mathcal P}_\text{\tiny LLL} {\mathcal H}_\text{220} {\mathcal P}_\text{\tiny LLL}$. However, that the first projection has no impact since the groundstate is already in the LLL and the second projection is irrelevant since the groundstate is the zero energy eigenstate of ${\mathcal H}_\text{eff}$.
\end{references}
\end{document}